%% file: paper.tex
\documentclass[acmlarge]{acmart}

\AtBeginDocument{%
  \providecommand\BibTeX{{%
    \normalfont B\kern-0.5em{\scshape i\kern-0.25em b}\kern-0.8em\TeX}}}

\setcopyright{acmcopyright}
\copyrightyear{2018}
\acmYear{2018}
\acmDOI{10.1145/1122445.1122456}

\acmJournal{POMACS}
\acmVolume{37}
\acmNumber{4}
\acmArticle{111}
\acmMonth{8}

\usepackage{graphicx}
\usepackage{subfig}
\usepackage{multirow}
\usepackage{url}
\usepackage{multicol}
\usepackage{epstopdf}
\usepackage{amsthm}
\usepackage{paralist}
\usepackage{listings}
\usepackage{balance}
\usepackage{grffile}

\usepackage{pgfplots}
\usetikzlibrary{intersections,patterns,trees,shapes,snakes}
\usetikzlibrary{fillbetween}
\usepackage{caption}
\usepackage{tikz}

\newtheoremstyle{theoremdd}
  {\topsep}
  {\topsep}
  {\itshape}
  {0pt}
  {\bfseries}
  {. ---}
  { }
  {\thmname{#1}\thmnumber{ #2}\thmnote{ (#3)}}

\theoremstyle{theoremdd}

\usepackage[linesnumbered,ruled,vlined]{algorithm2e}
\usepackage{balance}
\usepackage{resizegather}

\newcommand{\framework}{SDC-Scissor\xspace}
\newcommand{\ie}{\textit{i.e.,}\space}

\newcommand{\etc}{\textit{etc.}\space}
\newcommand{\etal}{\textit{et al.}\space}

\newcommand{\approach}{\textit{SDC-Prioritiz\-er}\xspace}
\newcommand{\mapproach}{\textit{MO-SDC-Prioritiz\-er}\xspace}
\newcommand{\sapproach}{\textit{SO-SDC-Prioritiz\-er}\xspace}
\newcommand{\apfdc}{\textit{APFD$_c$}\xspace}

\newcommand*\circled[1]{\tikz[baseline=(char.base)]{
            \node[fill=black, text=white,shape=circle,draw,inner sep=2pt] (char) {#1};}}

\newcommand{\finding}[2]{\vspace{2mm}
\noindent\fbox{\parbox{0.98\linewidth}{\selectfont
\textbf{Finding #1.} #2 }}\\}

\begin{document}
\newcommand\seba[1]{\textcolor{blue}{\nb{SEB: }{#1}}}
\newcommand\sajad[1]{\textcolor{purple}{\nb{SAJ: }{#1}}}
\newcommand\christian[1]{\textcolor{purple}{\nb{CHRIS: }{#1}}}
\newcommand\pouria[1]{\textcolor{orange}{\nb{Pouria: }{#1}}}
\newcommand{\revise}[1]{\textcolor{black}{{#1}}}
\revise{\title{Single and Multi-objective Test Cases Prioritization for Self-driving Cars in Virtual Environments}}

\author{Christian Birchler}
\affiliation{%
  \institution{Zurich University of Applied Science}
  \country{Switzerland}}
\email{birc@zhaw.ch}

\author{Sajad Khatiri}
\affiliation{%
  \institution{Zurich University of Applied Science \& Software Institute - USI, Lugano}
  \country{Switzerland}}
\email{mazr@zhaw.ch}

\author{Pouria Derakhshanfar}
\affiliation{%
  \institution{Delft University of Technology}
  \country{Netherlands}}
\email{p.derakhshanfar@tudelft.nl}

\author{Sebastiano Panichella}
\affiliation{%
  \institution{Zurich University of Applied Science}
  \country{Switzerland}}
\email{panc@zhaw.ch}

\author{Annibale Panichella}
\affiliation{%
  \institution{Delft University of Technology}
  \country{Netherlands}}
\email{a.panichella@tudelft.nl}


\begin{abstract}

\noindent 
\revise{Testing with simulation environments helps to identify critical failing scenarios for self-driving cars (SDCs). Simulation-based tests are safer than in-field operational tests and allow detecting software defects before deployment.} 
However, these tests are very expensive and are too many to be run frequently within limited time constraints.

\noindent
In this paper, we investigate test case prioritization techniques to increase the ability to detect SDC regression faults with virtual tests earlier.
\noindent 
\revise{Our approach, called \approach, prioritizes virtual tests for SDCs according to static features of the roads we designed to be used within the driving scenarios. These features can be collected without running the tests, which means that they do not require past execution results.} 
\revise{We introduce two evolutionary approaches to prioritize the test cases using diversity metrics (black-box heuristics) computed on these static features. 
These two approaches, called \sapproach{} and \mapproach{}, use single-objective and multi-objective genetic algorithms, respectively, to find trade-offs between executing the less expensive tests and the most diverse test cases earlier.}

\noindent
Our empirical study conducted in the SDC domain shows that 
\revise{
\mapproach{} significantly (p-value$<=0.1e-10$) improves the ability to detect safety-critical failures at the same level of execution time compared to baselines: random and greedy-based test case orderings.
Besides, our study indicates that multi-objective meta-heuristics outperform single-objective approaches when prioritizing simulation-based tests for SDCs.
}


\noindent 
\revise{
\mapproach prioritizes test cases with a large improvement in fault detection while its overhead (up to 0.45\% of the test execution cost) is negligible.
}

\end{abstract}



\maketitle

\section{Introduction}
\label{sec:introduction}

\newcommand{\rqone}{To what extent is it possible to prioritize safety-critical tests in SDCs in virtual environments prior to their execution?}
\newcommand{\rqtwo}{What is the cost-effectiveness of \approach compared to baseline approaches?} 
\newcommand{\rqtwoone}{What is the cost-effectiveness of \sapproach compared to baseline approaches?} 
\newcommand{\rqtwotwo}{What is the cost-effectiveness of \mapproach compared to baseline approaches?} 
\newcommand{\rqthree}{What is the overhead introduced by \approach?}
\newcommand{\rqthreeone}{What is the overhead introduced by \sapproach?}
\newcommand{\rqthreetwo}{What is the overhead introduced by \mapproach?}

Self-driving cars (SDCs) are autonomous systems that collect, analyze, and leverage sensor data from the surrounding environment to control physical actuators at run-time \cite{baheti2011cyber,national201721st}. 
%
Testing automation for SDCs is vital to ensure their safety and reliability \cite{KalraPaddock:2016,Kim2019}, but it presents several limitations and drawbacks:
\begin{inparaenum}[(i)]
\item the limited ability to repeat tests under the same conditions due to ever-changing environmental factors~\cite{Kim2019};
\item the difficulty to test the systems in safety-critical scenarios (to avoid irreversible damages caused by dreadful outcomes)~\cite{TheGuardian-2018,washingtonpost:2019,Ingrand19};
\item not being able to guarantee the system's reliability in its operational design domain due to a lack of testing under a wide range of execution conditions~\cite{KalraPaddock:2016}.
\end{inparaenum}

The usage of virtual simulation environments addresses several of the challenges above for SDCs testing practices
\cite{beamNG,BondiDKPSFDHIJT18,DosovitskiyRCLK17,nvidia_drive}. Hence, simulation environments are used in industry in multiple development stages of Cyber-physical Systems (CPSs)~\cite{Shin2018}, including model (MiL), software (SiL), and hardware in the loop (HiL). As a consequence, multiple open-source and commercial simulation environments have been developed for SDCs, which can be \textit{more effective and safer than traditional in-field testing methods} \cite{afzal2020study}. 

Adequate testing for SDCs requires writing (either manually or assisted by generation tools~\cite{Gambi2019,abdessalem2018testing}) a very large number of driving scenarios (test cases) to assess that the system behaves correctly in many possible critical and corner cases. The large running time of simulation-based tests and the large size of the test suites make regression testing particularly challenging for SDCs \cite{YohanandhanEMM20,Flores-GarciaKY20}. In particular, regression testing requires running the test suite before new software releases to assess that the applied software changes do not impact the behavior of the unchanged parts~\cite{Yoo:2010, DBLP:journals/tse/NucciPZL20}.


\revise{The \textit{\textbf{goal}} of this paper is to investigate and propose black-box test case prioritization (TCP) techniques for SDCs.}
TCP methods sort (prioritize) the test cases with the aim to run the fault-revealing tests as early as possible~\cite{Yoo:2010}. While various black-box heuristics have been proposed for traditional systems and CPSs, they cannot be applied to SDCs \textit{as is}. Black-box approaches for ``traditional'' systems sort the tests based on their diversity, computed on the values of the input parameters~\cite{ledru2012prioritizing} and the sequence of method calls~\cite{chen2018test}. However, SDC  simulation scenarios (e.g., with road shape, weather conditions) do not consist of sequences of method calls as in traditional tests~\cite{abdessalem2018testing, Gambi2019}. Approaches targeting CPSs measure test distance based on signal~\cite{arrieta2018multi}, and fault-detection capability~\cite{DBLP:journals/jss/ArrietaWSE19}. However, this data is unknown up-front without running the SDC tests. 

The main \textit{challenges} to address when designing black-box TCP methods for SDCs concern (i) the definition of features that can characterize SDC safety-critical scenarios in virtual tests; and (ii) design optimization algorithms that successfully prioritize the test cases based on the selected features.
Therefore, to address these challenges, we formulated the following \textit{research questions}:
\begin{itemize}
\item \revise{\textbf{RQ$_1$}: \textit{\rqone} }
\end{itemize}


\revise{We designed and computed 16 static features for driving scenarios in SDCs virtual tests, such as the length of the road, the number of left and right turns, etc. These features are extracted from the test scenarios prior to their execution, and for them, we investigated which ones are
  non-collinear (see Section \ref{sec:pca}) according to Principal Component Analysis (PCA).}
Hence, we introduce \approach, a TCP approach based on Genetic Algorithms (GA) that prioritizes test cases of SDCs by leveraging these features.
\revise{
This paper introduces two variants of the \approach, namely \sapproach and \mapproach. The former variant utilizes a single-objective genetic algorithm for test prioritization. The latter variant leverages a well-known and commonly used multi-objective genetic algorithm, called NSGA-II \cite{deb2002fast}, to achieve this goal. Any search-based technique needs to balance between \textit{exploitation} and \textit{exploration} \cite{vcrepinvsek2013exploration}. Exploitation refers to the ability of the search process to visit regions of the search space within the neighborhood of previously generated solutions (here, test execution orders). Exploration refers to the ability to generate entirely new solutions that are different from the current solutions. Poor exploration ability of the search process leads to low diversity between the generated solution, and thereby the search process may easily be trapped in local optima \cite{vcrepinvsek2013exploration}. The rationale behind introducing \mapproach beside the \sapproach is to avoid the lack of exploration ability in \approach. The NSGA-II algorithm, utilized in \mapproach, provides well-distributed Pareto fronts and thereby brings sufficient diversity into the generated solutions.}

\begin{itemize}
\item \textbf{RQ$_2$}:  \textit{\rqtwo}
\end{itemize}

\revise{To answer RQ$_2$, we conducted an empirical study with three different datasets and composed of test scenarios that target the \textit{lane-keeping} features of SDCs. In this context, fault-revealing tests are virtual test scenarios in which a self-driving car would not respect the \textit{lane tracking safety requirement} \cite{GambiMF19}.  
We targeted \texttt{BeamNG} by BeamNG.research~\cite{beamNG} (detailed in Section \ref{sec:rel-work}) as a reference simulation environment, which has been recently used in the Search-Based Software Testing (SBST) tool competition\footnote{https://sbst21.github.io/tools/}\cite{SBST2021}}.  
The test scenarios for this environment have been produced with by \framework \cite{Birchler:2022} (which integrates also AsFault \cite{Gambi2019}), an open-source project that generates test cases to assess SDCs behavior (detailed in Section \ref{sec:rel-work}).

By comparing \revise{\sapproach and \mapproach} with two baselines ---namely random search, and the greedy algorithm--- on these three benchmarks, we analyze the performance of our \revise{techniques} in terms of its ability to detect more faults while incurring a lower test execution cost.

Finally, we assess whether \approach \revise{techniques} can be used in practical settings, i.e., it does not add a too large computational overhead to the regression testing process:
\begin{itemize}
    \item  \textbf{RQ$_3$}:  \textit{\rqthree}  
\end{itemize}

The results of our empirical study show that \revise{\mapproach is the best performing technique in terms of identifying  more safety-critical scenarios in less time. On average, this technique reduces the time required to identify more safety-critical scenarios by 6\%, 25.5\%, and 3\% compared to \sapproach, random test case orders (``default'' baselines for search-based approaches~\cite{Shin2018, Yoo:2010}), and the greedy algorithm for TCP, respectively. It also shows that \revise{\mapproach} leads to an increase of detected faults (about 63 more) in the first 20\% of the test execution time compared to the greedy test prioritization (\ie second best technique according to our assessments).}
Furthermore, \approach \revise{approaches} do not introduce significant computational overhead in the SDCs simulation process, which is of critical importance to SDC development in industrial settings. 

The contributions of this paper are summarized as follows:
\begin{enumerate}
\item  \revise{We designed static features that can be used to characterize safe and unsafe test scenarios prior to their execution in the SDC domain}.
\item \revise{We introduce \sapproach and \mapproach, two black-box TCP approaches that leverage single and multi-objective Genetic algorithms, respectively, to achieve cost-effective regression testing with SDC tests in virtual environments.}
\item A comprehensive and publicly available replication package available on Zenodo \cite{anonymousRP}, including all data used to run the experiments as well as the prototype of \approach, to help other researchers reproduce the study results.
\end{enumerate}

\textbf{Paper Structure}.  
In Section \ref{sec:rel-work}, we summarize the related work, while in Section \ref{sec:approach}, we outline the approach we have designed and implemented to answer our research questions.
In Section \ref{sec:study}, we present our methodology and empirical studies performed to answer our research questions. 
In Section \ref{sec:results}, we report the study results, while in Section \ref{sec:threats}, we detail the threats to validity of our study.
Finally, Section \ref{sec:conclusions} concludes our study, outlining directions for future work.


\section{Background and Related Work}
\label{sec:rel-work}

This section discusses the literature concerning (i) test prioritization approaches in traditional systems; and (ii) studies closely related to test prioritization practices in the context of  CPSs (Cyber-physical systems). Finally, the section describes the background on the SDC virtual environment adopted in this study.

\subsection{Test Prioritization}
\label{sec:background:tp}
Approaches aiming at reducing the cost of regression testing can be classified into three main categories~\cite{Yoo:stvr2010}: \textit{test suite minimization}~\cite{Rothermel:icsm1998}, \textit{test case selection}~\cite{Chen:1996}, and \textit{test case prioritization}~\cite{Rothermel:icsm1999}. Test case minimization approaches tackle the regression problem by removing test cases that are redundant according to selected testing criteria (e.g., branch coverage). Test case selection aims to select a subset of the test suite according to the software changes, coverage criteria, and execution cost. Test case prioritization, which is the main focus of our paper, sorts the test cases to maximize some desired properties (e.g., code coverage, requirement coverage) that lead to detecting regression faults as early as possible. A complete overview of regression testing approaches can be found in the survey by Yoo and Harman~\cite{Yoo:stvr2010}.

\subsubsection{Prioritization heuristics} 
Approaches proposed in the literature to guide the prioritization of the test cases can be grouped into white-box and black-box heuristics~\cite{Yoo:stvr2010}. White-box test case prioritization uses past coverage data (e.g., branch, line, and function coverage) and iteratively selects the test cases that contribute to maximizing the chosen code coverage metrics. 

Black-box prioritization techniques rely on diversity metrics and prioritize the most diverse test cases within the test suites (e.g., ~\cite{feldt2016test,ledru2012prioritizing,Hajjaji4}). Widely-used diversity metrics include \textit{input and output set diameter}~\cite{feldt2016test}, or the \textit{Levenstein distance} computed on the input data~\cite{ledru2012prioritizing} and method sequence~\cite{chen2018test}. Further heuristics include topic modeling ~\cite{thomas2014static}, or models of the system~\cite{hemmati2013achieving}. Miranda \etal \cite{miranda2018fast} proposed fast methods to speed up the pair-wise distance computation, namely \textit{shingling} and \textit{locality-sensitive hashing}.
Recently, Henard \etal \cite{henard2016comparing} empirically compared many white-box and black-box prioritization techniques. Their results showed a large overlap between the regression faults that can be detected by the two categories of techniques and that
black-box techniques are highly recommended when the source code is not available ~\cite{henard2016comparing}, e.g., in the case of third-party components. Cyber-physical systems (including SDCs) are typical instances of systems with many third-party components~\cite{Shin2018}.

Prioritization heuristics for CPSs differ from those used for traditional software \cite{Arrieta2016AUTO}. We elaborate more in detail on the related work on test case prioritization for CPSs in Section \ref{sec:cps}.

\subsubsection{Optimization algorithms}
Given a set of heuristics (either white-box or black-box), optimization algorithms are applied to find a test case order that optimizes the chosen heuristics. As shown by Yoo \etal \cite{Yoo:stvr2010} test case prioritization (and regression testing in general) is inherently a multi-objective problem because test quality (e.g., code coverage, input diversity) and execution resources are conflicting in nature. The challenge is choosing balanced trade-offs that favor lower execution cost over higher code coverage or test diversity depending on the time constraints and resource availability (e.g., in  continuous delivery or integration servers).

Cost-cognizant greedy algorithms are well-known deterministic algorithms introduced for the set-cover problem and adapted to regression testing~\cite{Chen:1996}. The greedy algorithm first selects the test case with the most code coverage (white-box) or the most diverse one (black-box). Then, the algorithm iteratively selects the test case that increases coverage the most or that is the most diverse w.r.t. previously selected test cases~\cite{Yoo:stvr2010}. 

Meta-heuristics have been shown to be very competitive, sometimes outperforming greedy algorithms~\cite{DBLP:journals/tse/LiHH07, marchetto2015multi, DBLP:journals/tse/NucciPZL20, thomas2014static}. Marchetto \etal \cite{marchetto2015multi} used multi-objective genetic algorithms to optimize trade-offs between \textit{cumulative code coverage}, \textit{cumulative requirement coverage}, and \textit{execution cost}. 
Besides, genetic algorithms have been widely used to optimize test case diversity~\cite{thomas2014static} for black-box TCP. 

This paper uses \revise{greedy algorithm, single-objective genetic algorithm, and multi-objective genetic algorithm} to prioritize simulation-based test cases for self-driving cars. \revise{This is because each type of algorithm has been shown to outperform its counterparts in different domains and programs}~\cite{DBLP:journals/tse/LiHH07, DBLP:journals/jss/ArrietaWSE19}.

\subsection{Regression Testing for CPSs}
\label{sec:cps}
Regression testing is particularly critical for CPSs, which are characterized by interactions with simulation and hardware environments. Testing with simulation environments is a \textit{de facto} standard for CPSs, and it is typically performed at three different levels~\cite{matinnejad2013automated}: MiL, SiL, and HiL. During \textit{model in the loop} (MiL), the controller (cars) and the environments  (e.g., roads) are both represented by models, and testing aims to assess the correctness of the control algorithms. During \textit{software in the loop} (SiL), the controller model is replaced by its actual code (software), and its testing phase aims to assess the correctness of the software and its conformance to the model used in the MiL. Finally, during hardware in the loop  (HiL), the controller is fully deployed while the simulation is performed with real-time computers that simulate the physical signals. The testing phase for the HiL aims to assess the integration of hardware and software in more realistic environments~\cite{matinnejad2013automated}.

Regression testing for CPSs is more challenging as the execution time of the test cases is much longer due to the simulation~\cite{DBLP:journals/jss/ArrietaWSE19}. Hence, researchers have proposed different regression testing techniques that are specific to CPSs. Shin \etal \cite{DBLP:conf/issta/ShinNSBZ18} proposed a bi-objective approach based on genetic algorithms to prioritize acceptance tests for a satellite system. Their approach prioritizes the test cases according to the hardware damage risks it can expose (first objective) and maximizes the number of test cases that can be executed within a given time budget (second objective). 
Arrieta \etal \cite{DBLP:journals/jss/ArrietaWSE19} used both greedy algorithms and meta-heuristics to prioritize test cases for CPS product lines and with different test levels. 
In further studies, Arrieta \etal \cite{DBLP:journals/tii/ArrietaWMSE18} focused on multiple objectives to optimize for both test case generation and test case prioritization for CPSs. The objectives include requirement coverage, test case similarity, and test execution times. While test similarity for non-CPS systems is computed based on the lexicographic similarity for the method calls and test input, Arrieta \etal measured the similarity between the test cases based on the signal values for all the states in the simulation-based test case. 
Test case similarity computed at the signal-level has also been investigated in the context of test case selection for CPS~\cite{arrieta2018multi, DBLP:conf/splc/ArrietaWSE16}. 
\smallskip 

\textbf{Our paper} differs from the papers above w.r.t. the application domain and the optimization objectives. In particular, we focus on prioritized simulation-based test cases to assess the \textit{lane-keeping} features of self-driving cars. Instead, prior work focused on different domains, such as satellite~\cite{Shin2018}, electric windows~\cite{arrieta2018multi}, industrial tanks~\cite{DBLP:journals/jss/ArrietaWSE19, DBLP:journals/tii/ArrietaWMSE18}, and cruise controller~\cite{DBLP:journals/tii/ArrietaWMSE18}.
In our context, test cases consist of driving test scenarios with virtual roads (e.g., see Figure~\ref{fig:asfault_network}) and aim at assessing whether the simulated cars violate the lane-keeping requirements. 

Another important difference is related to the objectives (or heuristics) to optimize for regression testing. Prior works for CPS prioritize the test cases based on fault-detection capabilities~\cite{DBLP:journals/jss/ArrietaWSE19}, and diversity measured for simulation signals~\cite{DBLP:journals/tii/ArrietaWMSE18, arrieta2018multi, DBLP:conf/splc/ArrietaWSE16}. However, the fault-detection capability of the test cases is unknown a prior (i.e., without running the tests). Signal analysis requires knowing the states of the simulated objects in each simulated time step, which is also unknown before the actual simulation. Furthermore, a driving scenario (in our context) is not characterized by signals but only by the initial state of the car and the actual characteristics (e.g., shape) of the roads. Hence, we define features and diversity metrics that consider only the (static) characteristics of the roads that are used for the simulation. Unlike fault-detection capability and signals, our features can be derived from the driving scenario before the actual test execution.

\subsection{Background on SDCs Simulation}

\subsubsection{Main simulation approaches.}
Simulation environments have been developed to support developers in various stages of design and validation.
In the SDC domain, developers rely mainly on basic simulation models~\cite{10.1145/3239372.3239409,DBLP:journals/tits/SontgesA18}, rigid-body ~\cite{8877728,10.1007/978-3-030-60508-7_9}, and soft-body simulations~\cite{DBLP:conf/sigsoft/GambiHF19,RiccioTonella_FSE_2020}.

\emph{Basic simulation models}, such as MATLAB/Simulink models~\cite{10.1145/3239372.3239409,DBLP:journals/tits/SontgesA18}, implement fundamental signals but target mostly non-real-time executions and generally lack photo-realism. Consequently, while they are utilized for model-in-the-loop simulations and Hardware/Software co-design, they are rarely used for integration and system-level software testing.

\emph{Rigid-body simulations} 
approximate the physics of static bodies (or entities), i.e., by modeling them as \textit{undeformable} bodies. Basic simulation bodies consist of three-dimensional objects such as cylinders, boxes, and convex meshes \cite{abdessalem2018testing}. 

\emph{Soft-body simulations} can 
simulate deformable and breakable objects and fluids; hence, they can be used to model a wide range of simulation scenarios. Specifically, the finite element method (FEM) is the main approach for solid body simulations, while the finite volume method (FVM) and finite difference method (FDM) are the main strategies for simulating fluids~\cite{conf/wsc/MesitG11}.

\emph{Rigid-body v.s. Soft-body simulations}  Both rigid- and soft-body simulations can be effectively combined with powerful rendering engines to implement photo-realistic simulations~\cite{DosovitskiyRCLK17,BondiDKPSFDHIJT18,XuLXXYHMW19,beamNG}.
However, soft-body simulations can simulate a wider variety of physical phenomena compared to rigid-body simulations. 
Soft-body simulations are a better fit for implementing safety-critical scenarios (e.g., car incidents~\cite{DBLP:conf/sigsoft/GambiHF19}), in which a high simulation accuracy is of key importance. \revise{As follows, we describe the soft-body environment we used in our research investigation, i.e., BeamNG \cite{beamNG}.}

\revise{\subsubsection{BeamNG \& AsFault.} Creating adequate test scenario suites for SDCs is a hard and laborious task.
To tackle this issue, Gambi {\em et al.} \cite{GambiMF19} developed and proposed a tool called AsFault \cite{Gambi2019} to generate driving scenarios for testing SDCs automatically.  
From a high-level point of view, AsFault combines procedural content generation
and search-based testing
in order to automatically create virtual scenarios for testing the lane-keeping behavior in SDC software. 
Specifically, AsFault leverages a genetic algorithm to iteratively refine virtual road networks towards those which cause  the \textit{\textbf{ego-car}} (the simulated car controlled by the SDC software under test) to move away from the center of the lane.
The virtual roads are generated inside a driving simulator called BeamNG \cite{beamNG}, which can generate photo-realistic, but synthetic, images of roads. 
Given such characteristics, BeamNG \cite{beamNG} has also been used as the main simulation platform in the 2021 edition of the SBST tool competition~\cite{SBST2021}.
Lane-keeping systems (described in the next sections) continuously track the striped and solid lane markings of the road ahead using advanced image processing, deep learning, or machine learning techniques and triggers needed control mechanisms (e.g., steering, braking, and speeding) to keep the car at the proper location regarding the road structure.
}  

To evaluate the criticality of generated test cases, the road networks are instantiated in a driving simulation, during which the ego-car is instructed to reach a target location following a navigation path selected by AsFault.
During the simulation, AsFault traces the position of the ego-car at regular intervals such that it can identify Out of Bound Episodes (OBEs), i.e., lane departures. 
An \textbf{out-of-bound incident} is defined as \textit{``the case when the car went more than two meters out of the lane center"}. 
In our experiments, we use this information to label test scenarios as \textbf{\textit{safe}} (causing no OBEs) or \textbf{\textit{unsafe}} (causing at least one OBE).

\revise{Figure \ref{fig:asfault_network} illustrates a sample test scenario generated and executed by AsFault~\cite{GambiMF19}.
It includes start and target points for the ego-car on the map, the whole road network, the selected driving path (colored in yellow), and the detected OBE locations during the execution of the scenario by the ego-car.  
Hence, each generated test scenario by AsFault  
consists of a JSON file generated by AsFault, which reports multiple \textit{nodes} and their connections, and form a \textit{road network}, with the start and destination point and the driving path of the ego-car~\cite{GambiMF19}.
}



\subsubsection{SDC Software Use-cases}
\label{sec:AsFault-AIs}
\revise{AsFault supports two AI engines as test subjects while generating test cases, which we use to generate our test suites. These two test subjects allow to drive the ego-car by computing an ideal driving trajectory, which places the ego-car in the center of the lane while driving within a configurable speed limit}: 
\begin{itemize}
    \item \revise{\textbf{BeamNG.AI.} \footnote{\url{https://wiki.beamng.com/Enabling_AI_Controlled_Vehicles\#AI_Modes}}
        BeamNG.research ships with a driving AI that we refer to as \textit{BeamNG.AI}. 
        BeamNG.AI can be parameterized with an \textit{``aggression”} factor which controls the amount of \textit{risk} the driver takes in order to reach the destination faster.
        BeamNg.research developers say that low aggression factors (e.g., 0.7) result in a smooth driving whereas high aggression factors (e.g., 1.2 and above) lead the car to edgy driving and might cut corners \cite{GambiMF19}}. 
    
    \item \revise{\textbf{Driver.AI.}
    \footnote{\url{https://github.com/alessiogambi/AsFault/blob/asfault-deap/src/asfault/drivers.py}}
        Driver.AI is a trajectory planner shipped with AsFault~\cite{GambiMF19}.
        AsFault leverages an extension of Driver.AI, which monitors the quality of its predictions at run-time. Hence, differently from BeamNG.AI, Driver.AI analyzes the road geometry and plans the trajectory of the car by computing, for each turn, the maximum safe driving speed ($v$) using the reference formula for centripetal force on flat roads with static friction ($\mu$)~\cite{misc:safe-speed}:
    \begin{equation}
     v = \sqrt{\mu \times r \times g}
    \end{equation}
    where $r$ is the turn radius and $g$ is the free-fall acceleration. 
    It is important to note that, we use BeamNG since:
    \begin{itemize}
        \item BeamNG can be easily used by developers via Python APIs for creating scenarios
        \item BeamNG can access to sensor data, Camera, Lidar, IMU 
        \item the BeamNG AI engine can simulate:
        \begin{itemize}
            \item the aggressive driving style
            \item Balanced driving style
            \item Calm driving style
        \end{itemize}
    \end{itemize}}
\end{itemize}

\section{Approach}
\label{sec:approach}
This section describes the investigated test scenario features and prioritization strategies introduced by \approach and a greedy algorithm in the SDC domain. 

\subsection{SDC Road Features}
\label{sec:features}
 
In the context of SDC, we target the definition of features (or metrics) that characterize SDC tests in virtual environments according to the following requirements: the features (1) can be extracted before the actual execution of the virtual tests; and (2) these features can characterize (or identify) safe and unsafe scenarios without executing them. 
In the following, we describe how the SDC features have been designed and measured considering the BeamNG as the targeted SDC virtual environment.

\begin{figure}
    \centering
    \includegraphics[width=0.35\linewidth]{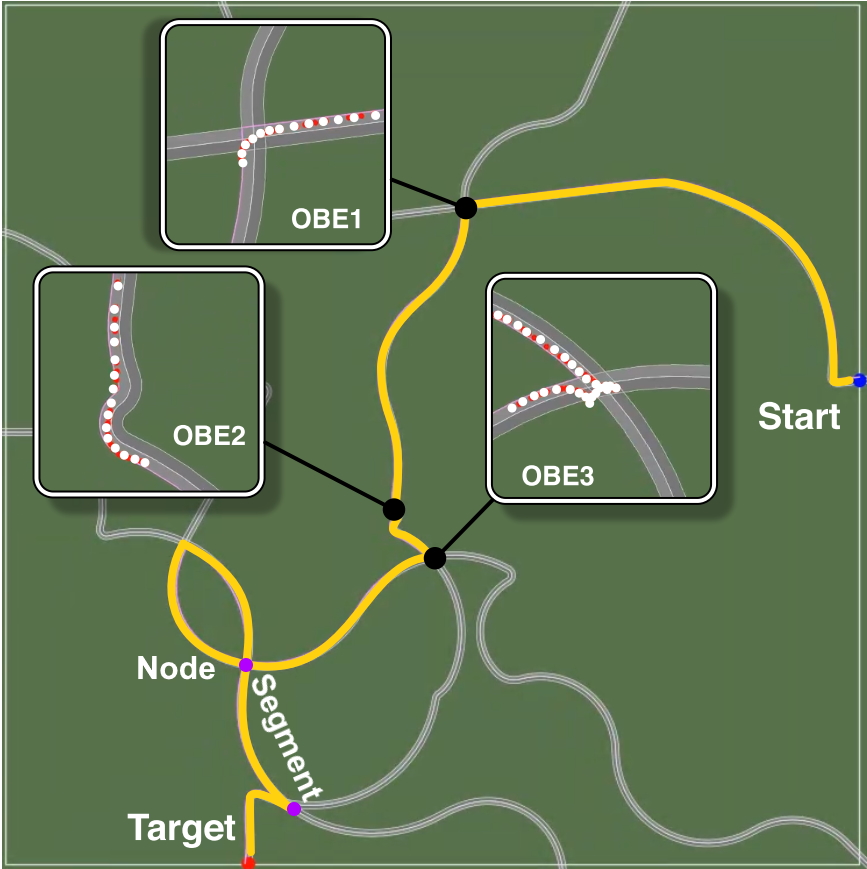}
    \caption{Sample driving scenarios generated by \framework \cite{Birchler:2022} (which integrates also AsFault \cite{GambiMF19}})
    \label{fig:asfault_network}
    \vspace{-4mm}
\end{figure}

In the context of BeamNG, it is possible to compute static features concerning the actual road characteristics of SDC virtual tests. 
Specifically, as illustrated in Figure \ref{fig:asfault_network}, each virtual test scenario generated by AsFault (virtual roads), consists of multiple \textit{nodes} and their connections (i.e.\textit{road segments}) forming a so-called \textit{road network}, along with the start and destination points and the driving path of the ego-car. 
This allows us to compute what we call \textit{Road Features}, i.e., features or characteristics of the road that will be used during the simulation within the BeamNG virtual environment. 

 From the road data reported by AsFault, we extract various features for each test scenario (as described in the following paragraph), and we investigate ways to leverage these features to determine the criticality of the test scenarios (as described in Section \ref{sec:study}).  

\textit{Road Features extraction.} To extract the features corresponding to each of the generated test scenarios, 
we leverage the JSON file generated as output by AsFault. 
These files, as explained before, consist of multiple \textit{nodes} and their connections, and form a \textit{road network}, with the start and destination point and the driving path of the ego-car.
Hence, we extract two sets of road features, \textit{the general road characteristics}, and \textit{the road segment statistics}.
The general road characteristics are attributes that refer to the road as a whole, e.g., direct distance and road length between the start and destination points, the total number of turns to left or right.
For each road segment (see figure \ref{fig:asfault_network}), we can extract individual metrics such as road angle and pivot radius. 
For the segment statistics features, we apply aggregation functions (e.g. minimum, maximum, average) on these individual segment metrics for all road segments in the scenario path. 
Table \ref{table:road_general_feat} reports the features extracted from the original fields in AsFault JSON (i.e., F1-16 features), specifying their description, type, and expected range of values for each feature. 
In the next sections, we described how the designed features are used as inputs to test case prioritization strategies.

\begin{table}
    \centering
    \caption{Road Characteristics Features}
    \label{table:road_general_feat}
    \small
    \begin{tabular}{ cllcc }
     \toprule
     \textbf{ID} & \textbf{Feature}  
      & \textbf{Description} & \textbf{Type} & \textbf{Range}\\
     \midrule
     F1 & Direct Distance 
       & Euclidean distance between start and finish & float & [0-490] \\
     \hline
     F2 & Road Distance 
       & Total length of the road & float & [56-3,318]\\
     \hline
     F3 & Num. Left Turns 
       & Number of left turns on the test track & int & [0-18]\\
     \hline
     F4 & Num. Right Turns 
       & Number of right turns on the test track & int & [0-17] \\
     \hline
     F5 & Num. Straight 
       & Number of straight segments on the test track & int & [0-11]   \\
     \hline
     F6 & Total Angle 
       & Total angle turned in road segments on the test track & int & [105-6,420] \\
     \hline
     F7 & Median Angle 
       & Median of angle turned in road segment on the test track & float & [30-330]   \\
     \hline
     F8 & Std Angle 
       & Standard deviation of angled turned in road segment on the test track & int & [0-150] \\
     \hline
     F9 & Max Angle 
       & The maximum angle turned in road segment on the test track & int & [60-345] \\
     \hline
     F10 & Min Angle 
       & The minimum angle turned in road segment on the test track & int & [15-285] \\
      \hline
     F11 & Mean Angle 
       & The average angle turned in road segment turned on the test track & float & [5-47] \\\hline
     F12 & Median Pivot Off 
       & 
     Median of radius of road segment on the test track & float & [7-47] \\ \hline
    F13 &  Std Pivot Off 
       & Standard deviation of radius of turned in road segment on the test track & float & [0-23] \\
     \hline
    F14 &  Max Pivot Off 
       & The maximum radius of road segment on the test track & int & [7-47] \\     \hline

    F15 &  Min Pivot Off 
       & The minimum radius of road segment on the test track & int & [2-47]  \\
     \hline
    F16 &  Mean Pivot Off 
       & The average radius of road segment turned on the test track & float & [7-47] \\     
     \bottomrule
    \end{tabular}
\end{table}

\subsection{
\revise{
Single-Objective Genetic Algorithm
}} 
\label{sec:ga}

\begin{figure}
    \centering
    \includegraphics[width=0.4\textwidth]{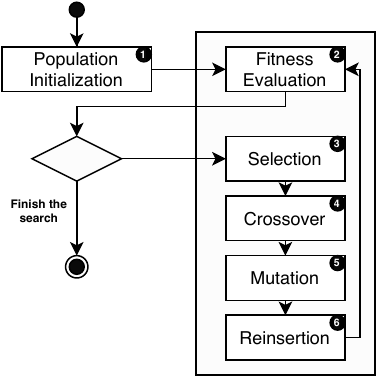}
    \caption{An overview of Genetic Algorithm}
    \label{fig:GA}
\end{figure}

Several prior studies have utilized evolutionary algorithms (particularly genetic algorithms) for test prioritization to reduce regression testing costs in different types of systems \cite{DBLP:journals/tse/LiHH07}. 
A typical Genetic algorithm (GA) starts with generating a population of randomly generated individuals (box \circled{1} in Figure \ref{fig:GA}). Each individual can be described as a sequence of parameters, called the chromosome, which encodes a potential solution to a given problem. This encoding can be performed in many forms (such as string, binary, \etc). After generating the first population, this algorithm determines the ``fitness'' of the individuals according to a fitness function (box \circled{2} in Figure \ref{fig:GA}). Then, in the \textit{Selection} phase (box~\circled{3} in Figure \ref{fig:GA}), a subset of individuals are selected according to their fitness values to be used as parents for mating. Next, two genetic operators are applied to generate the next population using the selected parents: \textit{Crossover} and \textit{Mutation}. The former (box \circled{4} in Figure \ref{fig:GA}) operator combines two parents to produce new individuals (called offspring). The latter (box \circled{5} in Figure \ref{fig:GA}) operator alters one or more elements in the offspring to explore nearby solutions in the search space. 
Finally, the newly generated individuals are saved in a new population (box \circled{6} in Figure \ref{fig:GA}).
The process of generating a new population of individuals from the previous one will continue until either the search objective is fulfilled or when the algorithm reaches the maximal number of generations (iterations).

\revise{This section introduces a single-objective genetic algorithm called \sapproach} that prioritizes the most diverse tests (according to their corresponding feature vectors) per unit of cost in self-driving cars.
\revise{
 The following subsections describe detailed information regarding the encoding, operators, and fitness function used in the \approach.}

\subsubsection{Encoding}
\label{sec:approach:encoding}
Since the solution for the test prioritization is an ordered sequence of tests, \approach uses a \textit{permutation encoding}.
\revise{
Assuming that, in our problem, we seek to order the execution of N tests, our approach encodes each chromosome as an N-sized array containing integers that denote the position of a test in the order. For example, let $\tau=\langle t_1, t_2, t_3 \rangle$ be a chromosome for a test suite with three test cases; then, test case $t_1$ will be executed first, followed by $t_2$ and $t_3$ during regression testing.
}

\subsubsection{Partially-Mapped Crossover (PMX)}
In the crossover, an offspring $o$ is formed from  two selected parents $p_1$ and $p_2$ , with the size of N, as follows: (i) select a random position $c$ in $p_1$ as the cut point; (ii) the first $c$ elements of $p_1$ are selected as the first $c$ elements of $o$; (iii) extract the $N - c$ elements in $p_2$ that are not in  $o$ yet and put them as the last $N - c$ elements of $o$.

\revise{\subsubsection{Mutation operators}
\label{sec:approach:mutation}
A chromosome $p$ can be mutated one or more times according to the given mutation probability.
In each round of mutation, one of the three following mutation operators\cite{10.1162/EVCO_a_00006} is selected randomly with an equal chance of ~0.33\% to perform the mutation:}
\revise{
\begin{itemize}
    \item \textbf{SWAP mutation:}
This mutation operator randomly selects two positions in a chromosome $p$ and swaps the index of two genes (test case indexes in the order) to generate a new offspring. 
    \item \textbf{INVERT mutation:}
This mutation operator randomly selects a segment (with a random size) of the given chromosome $p$. 
Then, it reverses the selected segment end to end and reattaches it to generate a new offspring.
    \item \textbf{INSERT mutation:}
This mutation randomly selects a gene in the chromosome $p$ and moves it to another index in the solution to generate a new offspring.
\end{itemize}
}

\revise{We consider the three operators above since prior studies~ \cite{10.1162/EVCO_a_00006} showed that using multiple mutation operators for permutation-based optimization problems increases the likelihood of escaping from solutions that are locally optimal under one mutation operator. This procedure used for the mutation is the same in both of the \approach variants introduced in this paper.
}

\subsubsection{Fitness function \revise {in \sapproach}} 
\label{sec:ff}
Our goal is to promote (1) the diversity of the selected test cases and (2) minimize the execution cost. \revise{Hence, the ultimate goal is to run the most diverse test within a given time constraint.} Hence, we define a fitness function that incorporates both test diversity and execution cost. 
\revise{
This is in line with current practice in the literature, which combines surrogate metrics for test effectiveness (e.g., code coverage) with execution cost~\cite{Yoo:2010zr,Li:2006,DBLP:journals/tse/NucciPZL20}.
}
More specifically, let $\tau = \langle t_1, \dots, t_n \rangle$ be a given test case ordering, its ``fitness'' (quality) is measured using the following equation:

\begin{eqnarray}
\label{eq:fitness}
    \max f(\tau) & = &\displaystyle \sum_{i=2}^{n} \frac{diversity(t_i)} {cost(t_i) \times i} \nonumber\\
    & = & \displaystyle \sum_{i=2}^{n} \frac{distance(t_i, t_{i-1})} {cost(t_i) \times i} 
\end{eqnarray}


where $n$ is the number of test cases; $t_i$ is the $i$-th test in the ordering $\tau$; $cost(t_i)$ is the execution cost (simulation time) of the test case $t_i$; and $distance(t_i, t_{i-1})$ measures the Euclidean distance between the test cases $t_i$ and $t_{i-1}$. In other words, each test case in position $i$ positively contributes to the overall fitness (to be maximized) based on its distance to the prior test $t_{i-1}$ in the order $\tau$. Since we want to have as many diverse tests as possible in the same amount of time, the diversity score of each test $t_i$ is divided by its execution cost (to be minimized) and its position $i$ in $\tau$. The factor $i$ in the denominator of Equation~\ref{eq:fitness} promotes solutions where test cases with the best diversity-cost ratio are prioritized early, i.e., they appear early within the order $\tau$. 

The distance between two tests $t_i$ and $t_j$ is measured using the Euclidean distance and computed on the feature vectors described in Table~\ref{table:road_general_feat}. It is important to highlight that the different features have different ranges and scales, as reported in Table~\ref{table:road_general_feat}. Hence, the distance values computed using the Euclidean distance might be biased toward the features with larger ranges. \revise{To remove this potential bias, we normalized the features using \textit{z-score} normalization, which is a well-known method to address outliers and to re-scale a set of features with different ranges and scales~\cite{geron2019hands}.} The \textit{z-score} normalization scale the features using the formula $\frac{x - \mu}{\sigma}$, where $x$ is the feature to re-scale, $\mu$ is its arithmetic mean, and $\sigma$ is the corresponding standard deviation~\cite{geron2019hands}.

The execution cost of each test case $t_i$ is estimated based on the past execution cost gathered from previous test runs, as recommended in the literature~\cite{elbaum2001incorporating, Yoo:2010}. This estimation is accurate for SDC since the cost of running simulation-based tests is proportional to the length of the road and the cost of rendering the simulation, which are fixed simulation elements.

\subsubsection{Selection \revise {in \sapproach}}
\revise{
The fitness function defined in Section~\ref{sec:ff} allows GAs to determine the fittest individual (permutations in our case) that should have higher chances to be selected for mating. 
}
The selection is made using the \textit{roulette wheel selection} \cite{goldberg2006genetic}, which assigns a selection probability to each of the individuals according to their fitness values (calculated by a fitness function). Assuming that our problem is a maximization problem, the selection probability of an individual $p_i$ is calculated as follows:
\begin{equation}
    \label{eq:selectionprob}
    P(p_i) = \frac{f(p_i)}{\sum_{j=1}^{N} f(p_j)}
\end{equation}
where $N$ is the number of individuals in the population and $f_i$ is the fitness value of $p_i$.

After allocating selection probability to individuals, the algorithm randomly selects some individuals according to their selection chance. Each individual with a lower fitness value has a lower allocated selection probability and thereby has a lower chance of transferring its genetic material to the next generation.

\subsection{
\revise{
Multi-objective Genetic Algorithm
}} 
\label{sec:approach:moff}
\revise{
This paper also proposes \mapproach{}, a multi-objective variant of \approach{} that considers the execution cost and test case diversity as two different objectives to optimize simultaneously. Assume that $\tau = \langle t_1, \dots, t_n \rangle$ is a solution (\ie test execution order) generated by the search process. The first goal to optimize is computed using the following equation:
}
\revise{
\begin{equation}
    \label{eq:f1-m}
    \max f_1(\tau) = \sum_{i=2}^{n} \frac{distance(t_i, t_{i-1})} {i}
\end{equation}
}
\revise{
\noindent where $distance(t_i, t_{i-1})$ denotes the distance between a test $t_i$ and its predecessor $t_(i-1)$ in the ordering. The contribution of each test case $t_i$ to the cumulative diversity is divided by its position $i$ in the ordering $\tau$. In other words, this objective promotes solutions where the most diverse test cases are executed earlier.
}

\revise{The second objective in \mapproach measures how steadily the cumulative cost increases when executing the tests with a given order $\tau$:
\begin{equation}
    \label{eq:f2-m}
    \min f_2(\tau) = \sum_{i=1}^{n} \frac{cost(t_i))} {i}
\end{equation}
\noindent where $cost(t_i)$ denotes the cost of executing the test case $t_i$ in $\tau$. The contribution of each test case $t_i$ to the cumulative cost is divided by its position $i$ in the ordering $\tau$, with the goal of promoting solutions where the least expensive test cases are executed earlier. Notice that this objective should be minimized.
}

\revise{
Different from \sapproach{}, finding optimal solutions for problems with multiple criteria requires trade-off analysis. Given the conflicting nature of our two objective\footnote{Diverse tests are not necessarily the least expensive to run}, it is not possible to obtain one single solution that optimizes both objectives at the same time~\cite{Coello:2006}. Hence, we are interested in finding the set of solutions that are optimal compromises between the two objectives.
}
\revise{
For multi-objective problems, the concept of optimality is based on concepts of \textit{Pareto dominance} and \textit{Pareto optimality}\cite{Coello:2006}.
In particular, a solution $\tau_A$ \textit{dominates} another solution $\tau_B$ ($\tau_A <_{p} \tau_B$) if and only if at the same level of diversity, $\tau_A$ has a lower cost than $\tau_B$. Alternatively, $\tau_A$ \textit{dominates} $\tau_B$ if and only if, at the same level of cost, $\tau_A$ has a larger diversity than $\tau_B$.
Among all possible solutions, we are interested in finding those that are not dominated by any other possible solution (\textit{Pareto optimality}). Pareto optimal solutions form the so-called \textit{Pareto optimal set} while the corresponding objective values form the \textit{Pareto front}.
}

\revise{
Figure~\ref{fig:dominance} provides a graphical example of Pareto optimality and non-dominance. All solutions in the grey rectangle (including $B$) dominate $D$ since they achieve both lower cost and higher diversity. Instead, all solutions in the blue rectangle (including $D$ and $E$) are dominated by $B$, since $B$ achieves higher diversity with lower execution cost. Finally, $A$, $B$, and $C$ do not dominate one another while $D$ and $E$ are dominated solutions.
}

\begin{figure}[t]
\centering
\pgfplotsset{width=0.5\linewidth} 
\input{graphs/dominance}
\caption{Graphical representation of Pareto dominance for our two objectives, namely (1) test diversity (to maximize) and test cost (to minimize). In the example, points $A$, $B$, and $C$ do not dominate one another, while point $B$ dominates both $D$ end $E$.}
\label{fig:dominance}
\end{figure}
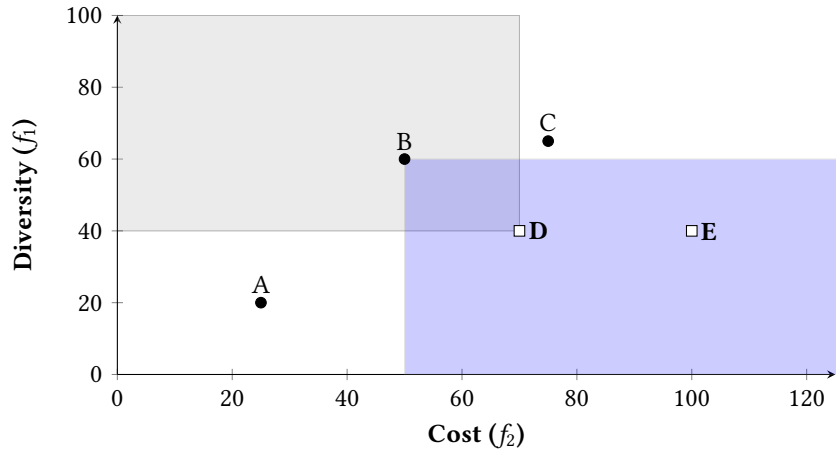

\revise{
\subsubsection{NSGA-II}
\label{sec:approach:moselection}
To find Pareto optimal solutions, \mapproach uses NSGA-II \cite{deb2002fast}.  This genetic algorithm provides well-distributed Pareto fronts and performs best when dealing with two or three search objectives \cite{deb2002fast}. NSGA-II shares the main loop of the genetic algorithm depicted in Figure~\ref{fig:GA}. Thus, it shares the same encoding schema as well as mutation and crossover operators discussed in Section~\ref{sec:ga}. However, it differs on how parents are selected for reproduction and how the new population is formed for the next generation. Parents are selected using the \textit{binary tournament selection}, which compares pairs of solutions in tournaments and selects the ``fittest'' solution from each pair for reproduction. Finally, the population for the next generation is obtained by selecting the ``fittest`` solutions among parent and offspring solutions (elitism).
}

\revise{
In NGSA-II, the ``fitness`` of the solutions is determined using the \textit{fast non-dominated sorting} algorithm and the concept of \textit{crowding distance}~\cite{Deb:2000}.
The former ranks the solutions according to their dominance relations. All non-dominated solutions within a given population are inserted in the first front $F_1$ (rank $r=1$); the subsequent front $F_2$ (rank $r=2$) contains all solutions that are dominated only by the solutions in $F_1$; and so on. Hence, solutions in the fronts with lower rank are ``fitter'' according to the Pareto optimality. 
}

\revise{
Instead, the crowding distance aims at promoting more diverse (isolated) solutions within each dominance rank. The crowding distance for a given solution is computed as the sum of the distances between such an individual and all the other individuals with the same rank. This heuristics is put in place to avoid selecting individuals that are too similar to each other.
}


\revise{
\subsubsection{Choosing a Pareto optimal solution}
\label{sec:approach:select:solution}
As explained in Section \ref{sec:approach:moselection}, NSGA-II returns a set of non-dominated solutions at the end of the search process. Hence, the next step is to decide which Pareto optimal solution (best trade-off) among the many different alternatives.
The necessity of this decision-making approach is also experienced in other optimization methods for various engineering problems~\cite{marler2004Survey}.
Researchers have suggested considering various points of interest in the Pareto front, such as the \textit{knee points}~\cite{branke:2004}, \textit{mid
points}~\cite{nagy2012lorenz}, or the \textit{extreme} of the Pareto front~\cite{panichella:2015}.
}

\revise{
One of the common techniques to select solutions from the Pareto front is to identify knee points~\cite{branke:2004,messaoudi2018search}, which are the solutions that minimize the distance to a point in the vector of the objective function, called \textit{Utopia Point}~\cite{marler2004Survey}. The utopia point is a (usually unreachable) point with the most-optimum observed value for each objective function. Assume that \mapproach returns a set of solutions $S={S_1, S_2, ...,S_i}$ as the final answer. These solutions are non-dominated according to two search objective functions diversity ($f_1(\tau)$ in Equation \ref{eq:f1-m}) and test execution cost ($f_2(\tau)$ in Equation \ref{eq:f2-m}). In this case, the Utopia Point $U$ is the following point in the two-dimensional objective functions vector:
\begin{equation}
    \label{eq:up-m}
    U=(maximum(\{f_1(s) | s \in S\}) , maximum(\{f_2(s) | s \in S\})
\end{equation}
Since the utopia point usually does not exist in the returned solutions, we select the \textit{closest} non-dominated solution to this point as the trad-off to select for regression testing.
}

\revise{
One common way to measure the distance between two points is using the \textit{Euclidean distance} $N(x)$, which is defined as:
\begin{equation}
    \label{eq:eucl-dist}
    N(x) = \sqrt{\sum_{i=1}^{k} (f_i(x) - U_i)^2}
\end{equation}
\noindent were $f_i$ is the value of the Pareto optimal solution $x$ for each  objective. Here, \mapproach has $f_1$ and $f_2$, as explained in Section~\ref{sec:approach:moff}. $U_i$ is also the value of the utopia point for the $i$th objective fitness function.
}

\revise{
It is notable that if the fitness functions have different units, the Euclidean norm becomes insufficient to represent the \textit{closeness}~\cite{marler2004Survey}. This is the case in \mapproach as the execution cost and the test diversity have different units. To tackle this issue, we need to normalize the values to make them dimensionless. The most robust technique to perform this normalization is~\cite{marler2004Survey, koski1987norm,rao1991modified}:
\begin{equation}
    \label{eq:norm}
    norm(f_i(x)) = \frac{f_i(x)- U_i}{max(f_i) - U_i}
\end{equation}
Where $f_i(x)$ is the fitness actual value of solution $x$ according to search objective fitness function $f_i$, and $max(f_i)$ is the maximum fitness value of generated solutions for $f_i$.
}

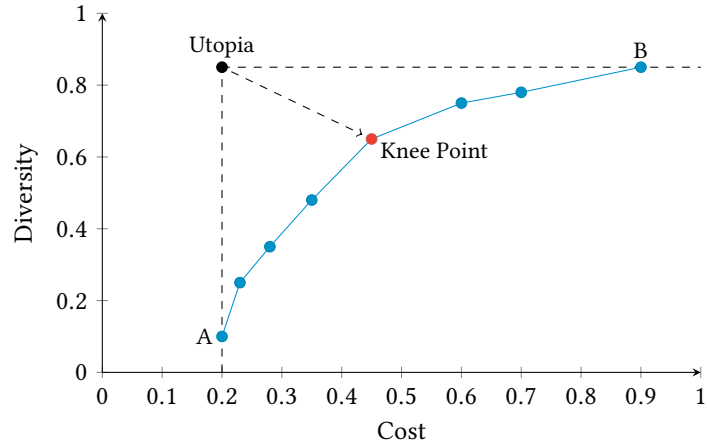
\begin{figure}[t]
\centering
\pgfplotsset{width=0.5\linewidth} 
\input{graphs/knee-point}
\caption{Graphical representation of a Pareto front (in blue), the utopia (black point), and the knee point (red point).}
\label{fig:dominance}
\end{figure}

\revise{
}

\subsection{Black-box Greedy Algorithm}
Greedy algorithms are well-known deterministic algorithms that iteratively build a solution (tests ordering) based on greedy steps. Greedy algorithms have been widely used in regression testing for both white- and black-box test case prioritization~\cite{Yoo:2010, thomas2014static}. Hence, we adapt the greedy algorithm to our context and use the set of features we have designed for SDCs (see Section~\ref{sec:features}).

The greedy algorithm first computes the pairwise distance among all test cases in the given test suite. Similarly to GAs, the distance between two test cases $t_i$ and $t_j$ is computed using the Euclidean distance between the corresponding feature vectors. These features are normalized up-front using the \textit{z-score} normalization as done for GA as well. 
\revise{
Then, the greedy algorithm computes the diversity per unit cost of each test $t_i$ using the following equation:
\begin{equation}
    \label{eq:selectionprob-m}
    score (t_i) = \frac{distance(t_i, \tau)} {cost(t_i))} 
\end{equation}
\noindent where $distance(t_i, \tau)$ measures the distance between $t_i$ and the tests $t_j \in \tau$ selected in the previous iterations of the algorithm. In this equation, a higher score for a test means that it has the highest dissimilarity to previously selected tests with the lowest execution cost. 
}

\revise{
The greedy algorithm initializes the test order $\tau$ by selecting the test with the largest ratio between (1) its average distance to all other tests in the suite and (2) its execution cost.
Then, the algorithm iteratively finds the test case (among the non-selected ones) with the largest average (mean) \revise{score} to the (already selected) test cases in $\tau$. This selection step corresponds to the \textit{greedy} heuristic. Suppose multiple tests have the same average \revise{score} to $\tau$. In that case, the tie is broken by randomly choosing one of the equally distant test cases. This process is repeated until all test cases are prioritized. 
}

\section{Study Design}
\label{sec:study}

\textbf{Study design overview}. Our empirical study is steered by the following research questions:
\begin{itemize}
    \item \revise{\textbf{RQ$_1$}: \textit{\rqone} }
    \item \textbf{RQ$_2$}: \textit{\rqtwo} 
    \item \textbf{RQ$_3$}: \textit{\rqthree}
\end{itemize}

In Section~\ref{sec:features}, we have introduced multiple static features to virtual driving scenarios (see Table \ref{table:road_general_feat}), some of which might be collinear or not useful for prioritizing test cases in a cost-effective way. 
Hence, our first research question (\textbf{RQ$_1$}) aims to determine which features to consider, by leveraging statistical methods based on collinearity analysis \cite{dormann2013collinearity,5601833}. 
Our second research question (\textbf{RQ$_2$}) aims to assess the extent to which test case orders produced by \approach \revise{techniques (\sapproach and \mapproach)}  can detect more faults (effectiveness) and with lower execution cost (efficiency) with respect to a naive random search. Specifically, as elaborated in detail later, a random search is a critical baseline for search-based solutions since it is a ``sanity-check'' to assess whether more ``sophisticated'' techniques are needed for a given domain~\cite{Shin2018}. In RQ$_2$, we compare the internal search algorithms discussed in Section~\ref{sec:approach}, namely the greedy algorithm, \revise{single-objective} \revise{and multi-objective genetic-algorithms}. 
With our last research question (\textbf{RQ$_3$}), we want to measure the overhead required to prioritize SDC test cases in virtual environments with \approach \revise{techniques}. This is an important aspect to investigate since a critical constraint in regression testing is that the cost of prioritizing test cases should be smaller than the time needed to run the test suite \cite{Yoo:2010}. 
Therefore, fast approaches are fundamental from a practical point of view to enable rapid and continuous test iterations during SDC development \cite{miranda2018fast}.

\subsection{Benchmark Datasets}
The benchmark used in our study consists of three experiments performed on corresponding datasets.
\revise{For each experiment, virtual test scenarios are generated and labeled as safe or unsafe by \framework \cite{Birchler:2022} (which integrates also AsFault).
As described in Table \ref{tab:datasets}, the first experiment leverages a dataset (referred to as \textit{BeamNG.AI.AF1}) that includes 1,178 virtual test scenarios generated with respect to BeamNG.AI with an aggression factor set to 1.} 
Since this is a cautious driving setup for BeamNG.AI, this dataset includes mostly safe scenarios, with about 26\% of the scenarios being unsafe (causing OBEs). 
For the second experiment, we created a new dataset (referred to as \textit{BeamNG.AI.AF1.5}) where we configured BeamNG.AI to drive in a more aggressive driving style.
\revise{This resulted in 5,638 test scenarios among which 45\% are unsafe.}

To increase the level of reliability and applicability of our results, we used another SDC  driving AI, namely Driver.AI, to generate the dataset of our last experiment. 
This last experiment was needed because using test scenarios with Driver.AI allows drawing a direct comparison with BeamNG.AI and investigating if the features we investigate are limited to BeamNG.AI or can be applied to other driving AIs.
Thus, we used \framework \cite{Birchler:2022} (which integrates also AsFault) to re-run the test scenarios in \textit{BeamNG.AI.AF1.5} 
with Driver.AI, resulting in a more cautious driving with only 19\% of the scenarios being unsafe.  

\begin{table}[]
\caption{ Datasets composition}
    \vspace{-1mm}
\label{tab:datasets}
\begin{tabular}{c|l|l|l|c}
\hline
\multirow{2}{*}{\textbf{Dataset}}        & \multicolumn{3}{c|}{\textbf{Number of Test Scenarios}} & \textbf{Running}  \\ \cline{2-4} 
                                         & \textbf{Unsafe} & \textbf{Safe} & \textbf{Total} & \textbf{Time}\\ \hline
\multicolumn{1}{l|}{BeamNG.AI.AF1}   & 312 (26\%)      & 866 (74\%)    & 1,178    & 16h      \\ \hline
\multicolumn{1}{l|}{BeamNG.AI.AF1.5} & 2,543 (45\%)    & 3,095 (55\%)  & 5,638    & 28h      \\ \hline
Driver.AI                              & 1,045 (19\%)    & 4,585 (81\%)  & 5,630   & 106h      \\ \hline
\end{tabular}
    \vspace{-2mm}
\end{table}

\subsection{Analysis Method}

\subsubsection{RQ$_1$: Feature Analysis}\label{sec:pca}
In a real scenario, we do not determine the tests' safety without executing all of them. Hence, we do not include the feature that indicates if a test is safe or unsafe in this research question. So, to answer our first research question, we analyze the orthogonality of the other 16 different features introduced in Section~\ref{sec:features}. In particular, we use the PCA to statistically assess whether all features are useful for test case prioritization or whether certain features are \textit{multicollinear}. A group of features is said to be \textit{collinear} if they are linearly related and implicit measures of the same phenomenon (road characteristics in our case). Addressing data collinearity is vital to avoid distance measurements being skewed toward the collinear features~\cite{dormann2013collinearity}. Besides, distance metrics (including the Euclidean distance) might not truly represent the extent to which the data points (test cases) are truly diverse when using a large number of features~\cite{du2016study}.

PCA is a well-founded, analytical, and established technique that allows to identify the orthogonal dimensions (principal components) in the data and measure the contributions of the different features to such components. Features that contribute to the same principal components are collinear and can be removed via dimensionality reduction. In particular, the PCA decomposes each dataset $M$ (e.g., \texttt{BeamNG.AI.AF1}) in two matrices:
$    \underset{m \times n}{M} \approx \underset{m \times n}{S} * \underset{n \times k}{V}$
In this equation, $m$ is the number of test cases; $n$ is the number of original features; $k$ is the number of principal components; $S$ denotes the features-to-component score matrix. More specifically, $S$ contains each feature's scores (contributions) to the latent components identified by the PCA. In an ideal dataset with zero collinearity, the features should exclusively contribute to different principal components. 

PCA can be used not only to detect but also to alleviate collinearity via dimensionality reduction~\cite{du2016study}. In particular, a lower-dimensional matrix can be obtained by choosing the top $h<k$ principal components and reconstructing the matrix as:
\begin{equation}
\label{eq:pca}
    \underset{m \times h}{M'} \approx \underset{m \times n}{S} * \underset{n \times h}{V}
\end{equation} 
Notice that $M'$ will contain new (non-collinear) features that are built as a combination of the old ones. This process is widely known in machine learning as \textit{feature extraction}~\cite{geron2019hands}. 

To answer RQ$_1$, we use PCA to detect (eventual) multi-collinearity among the different road features. In the case multi-collinearity is detected, we use PCA for \textit{dimensionality reduction} and \textit{feature extraction} by selecting the top $k$ principal components corresponding to 98\% of the original data variance, as recommended in the literature~\cite{geron2019hands}. 
The selected, relevant features in RQ$_1$ (discussed in Section \ref{sec:results:rq1}) are then considered to investigate RQ$_2$ and RQ$_3$ 
\revise{
and applied for all search algorithms, i.e., for both greedy and evolutionary algorithms.
}

\subsubsection{RQ$_2$: Cost-effectiveness of \approach Compared to Baseline Approaches}\label{sec:cos-effectivenes}
To assess the effectiveness of test case prioritization techniques introduced in this study, we look at the rate of fault detection (\ie how fast faults are detected during the test execution process). Hence, a better technique provides a test execution order that detects more faults while executing fewer tests. To indicate the rate of fault detection in our evaluation, we use a well-known metric in test case prioritization,  called Cost cognizant Average Percentage of Fault Detection (\apfdc)~\cite{malishevsky2006cost,elbaum2001incorporating,epitropakis2015empirical,jiang2009adaptive,rothermel1999test}. In this metric, higher \apfdc means a higher fault detection rate. 
Since there is no technique introduced for measuring the fault severity in the SDC domain, we consider the same severity for all of the faults. Hence, in our case, \apfdc can be formally defined as follows:
\begin{equation}
APFD_c=\frac{\sum_{i=1}^{m}\left(\sum_{j=T F_{i}}^{n} t_{j}-\frac{1}{2} t_{T F_{i}}\right)}{\sum_{j=1}^{n} t_{j} \times m}
\end{equation}
where $T$ is the list of tests that need to be sorted for execution; $t_j$ is the execution time required to run the test positioned as the $j$th test; $n$ and $m$ are the number of tests and faults, respectively; and $TF_i$ is the position in the given test permutation that detect fault~$i$. 
\revise{
We also assessed whether there is no significant variation in execution time (simulation time) of the simulation-based tests by executing them multiple times. In particular, we randomly selected 50 tests from our dataset and ran them ten times each. As a result, the average standard deviation of test execution time is 1.67s (less than 1\% variation) and the average coefficient of variance is 0.01.
}

To draw a statistical comparison between \revise{\sapproach}, \revise{\mapproach}, random search, and greedy algorithm, we use Vargha-Delaney $\hat{A}_{12}$ statistic \cite{vargha2000critique} to assess the effect size of differences between the \apfdc values achieved by these approaches. A value  $\hat{A}_{12}>0.5$ for a pair of factors (A, B) confirms that A has a higher fault detection rate and vise versa. Furthermore, to examine if the differences are statistically significant, we use the non-parametric Wilcoxon Rank Sum test, with  $\alpha = 0.05$ for Type I error.


\subsubsection{RQ$_3$: Overhead Introduced by \approach}
For RQ$_3$, we monitor the running time needed by \revise{\sapproach, \mapproach, } and the greedy algorithm to prioritize the test cases. This analysis aims to verify whether the extra overhead introduced by \approach \revise{techniques}, on average, leads to a disruption in the testing process or is negligible compared to the total time needed to run the entire test suite. To have a more reliable estimation of the running time, we run \revise{both \sapproach and \mapproach} 30 times and using the parameter values discussed in Section~\ref{sec:parameters}. Then, we measure the overhead of the different algorithms as the average running time over the 30 runs.

\subsubsection{Parameter setting}
\label{sec:parameters}
We used the default parameter values of the genetic algorithm as used in previous
studies on TCP (e.g.,~\cite{thomas2014static, DBLP:journals/tse/NucciPZL20, epitropakis2015empirical}). In particular, we use the following
parameter values:
\begin{itemize}
    \item \textit{Population size}: we used a pool of 100 test permutations.
    \item \textit{Crossover operator}: we used the partially-mapped crossover (PMX) for permutation problems (see Section \ref{sec:ga}) with a crossover probability $p_c$ = 0.80. This corresponds to the default value in Matlab and it is inline with the recommended range $0.45\leq p_c \leq 0.95$~\cite{cobb1993genetic, briand2006using}.
    \item \textit{Mutation operator}: we used the \revise{hybrid mutation operator, introduced in Section \ref{sec:approach:mutation},}  with a mutation probability $p_m$ = 1/$n$, where $n$ is the number of the test cases to prioritize. This choice is in line with the recommendations from previous studies~\cite{briand2006using,schaffer1989study} that showed how $p_m$ values proportional to the chromosome length produce better results.
    \item \revise{\textit{Stopping criterion}: the search ends after 4,000 generations (or equivalently 400K fitness evaluations).} We opted for a larger number of generations compared to prior studies in test case prioritization (e.g., \cite{DBLP:journals/tse/LiHH07,di2018test,DBLP:journals/jss/ArrietaWSE19}) since the test suites in our benchmark are much larger than those used in prior studies in TCP for traditional software (e.g., the programs in the SIR dataset~\cite{SIR}).
\end{itemize}
Notice that we did not fine-tune the parameters but opted for the default values. This choice is motivated by recent studies that showed that default values are a
reasonable choice in search-based software engineering~\cite{arcuri2013parameter, sayyad2013parameter}. Indeed, parameter tuning is a quite laborious and expensive process that does not assure better
performances when using meta-heuristics. Our initial experiments confirm this finding as default values already provide good results in our case.

\begin{table*}[t]
    \centering
    \caption{Results of the Principal Component Analysis (PCA) for BeamNG.AI.AF1. Values in boldface indicate the features that 
contribute the most to the main components (Cs) extracted by PCA.}
\vspace{-1mm}
\label{tab:pca:beamng1}
\revise{
\resizebox{.95\textwidth}{!}{
    \begin{tabular}{c | rrrrrrrrrrrrrrrr}
\hline
Features  &  C1 & C2 & C3 & C4 & C5 & C6 & C7 & C8 & C9 & C10 & C11 & C12 & C13 & C14 & C15 & C16\\
\hline
Direct Distance & -0.331 & 0.249 & \textbf{0.874} & -0.068 & 0.218 & -0.068 & -0.074 & -0.032 & 0.037 & -0.003 & -0.004 & -0.001 & 0.003 & 0.017 & 0.007 & -0.001 \\
Road Distance & 0.223 & -0.129 & 0.187 & 0.315 & 0.013 & -0.049 & \textbf{0.810} & -0.148 & 0.275 & -0.008 & 0.106 & -0.019 & 0.000 & 0.179 & 0.004 & 0.002 \\
Num. Left Turns & \textbf{0.421} & -0.033 & 0.189 & 0.127 & -0.098 & -0.213 & -0.137 & 0.014 & -0.233 & 0.040 & 0.063 & -0.224 & 0.002 & -0.076 & -0.329 & -0.687 \\
Num. Right Turns & 0.242 & -0.333 & 0.199 & 0.103 & -0.113 & -0.193 & -0.131 & -0.003 & -0.171 & -0.113 & -0.138 & \textbf{0.580} & 0.072 & 0.008 & \textbf{0.552} & -0.075 \\
Num. Straight & 0.196 & -0.144 & -0.121 & 0.059 & \textbf{0.956} & -0.012 & -0.059 & 0.002 & -0.063 & -0.002 & -0.005 & 0.011 & 0.013 & -0.031 & 0.004 & 0.002 \\
Total Angle & 0.388 & \textbf{0.797} & -0.089 & -0.016 & 0.034 & 0.016 & 0.035 & 0.041 & 0.084 & -0.318 & -0.080 & 0.293 & 0.028 & 0.028 & -0.005 & 0.006 \\
Median Angle & \textbf{0.437} & -0.041 & 0.200 & 0.134 & -0.107 & -0.224 & -0.147 & 0.018 & -0.243 & 0.063 & 0.067 & -0.217 & 0.013 & -0.086 & -0.169 & \textbf{0.718} \\
Std Angle & 0.151 & 0.299 & -0.034 & -0.021 & 0.012 & 0.020 & 0.008 & -0.059 & -0.046 & \textbf{0.534} & 0.092 & -0.384 & -0.034 & 0.006 & \textbf{0.656} & -0.081 \\
Max Angle & 0.119 & -0.073 & 0.025 & 0.199 & -0.020 & 0.090 & -0.448 & 0.015 & \textbf{0.419} & -0.036 & 0.235 & -0.058 & 0.056 & \textbf{0.702} & 0.005 & 0.002 \\
Min Angle & 0.147 & -0.007 & 0.055 & 0.122 & -0.021 & 0.126 & -0.170 & -0.119 & \textbf{0.484} & \textbf{0.556} & -0.089 & 0.367 & -0.155 & -0.372 & -0.230 & 0.015 \\
Mean Angle & -0.065 & 0.107 & -0.056 & -0.163 & 0.026 & -0.069 & 0.157 & -0.041 & -0.425 & \textbf{0.512} & -0.085 & 0.344 & 0.150 & 0.508 & -0.273 & 0.024 \\
Median Pivot Off & -0.273 & 0.159 & -0.121 & \textbf{0.656} & 0.007 & -0.076 & -0.115 & -0.414 & -0.246 & -0.053 & -0.170 & 0.004 & -0.412 & 0.060 & 0.002 & 0.004 \\
Std Pivot Off & -0.234 & 0.127 & -0.089 & \textbf{0.486} & 0.005 & -0.091 & -0.035 & 0.168 & -0.013 & 0.084 & 0.225 & 0.057 & \textbf{0.736} & -0.212 & -0.003 & -0.009 \\
Max Pivot Off & 0.061 & -0.009 & 0.111 & 0.096 & -0.017 & \textbf{0.633} & 0.028 & 0.085 & -0.328 & -0.023 & \textbf{0.610} & 0.196 & -0.198 & -0.085 & 0.002 & -0.002 \\
Min Pivot Off & 0.029 & -0.018 & 0.096 & 0.300 & -0.009 & 0.322 & 0.060 & \textbf{0.698} & -0.052 & 0.078 & -0.516 & -0.104 & -0.108 & 0.094 & 0.006 & -0.001 \\
Mean Pivot Off & -0.165 & 0.072 & -0.109 & 0.033 & 0.036 & -0.559 & 0.054 & 0.512 & 0.081 & 0.079 & 0.399 & 0.151 & -0.426 & -0.005 & 0.013 & -0.003 \\
\hline
Importance & 31.35\% & 25.72\% & 13.76\% & 9.00\% & 6.14\% & 3.96\% & 3.32\% & 2.14\% & 1.71\% & 1.10\% & 0.54\% & 0.48\% & 0.41\% & 0.22\% & 0.12\% & 0.01\%\\
\hline
    \end{tabular}
    }
    }
\end{table*}

\begin{table*}[t]
\centering
\caption{Results of the Principal Component Analysis (PCA) for BeamNG.AI.AF1.5. Values in boldface indicate the features that 
contribute the most to the main components (Cs) extracted by PCA.}
\vspace{-1mm}
\label{tab:pca:beamng15}
\revise{
\resizebox{.95\textwidth}{!}{
    \begin{tabular}{c | rrrrrrrrrrrrrrrr}
\hline
Features  &  C1 & C2 & C3 & C4 & C5 & C6 & C7 & C8 & C9 & C10 & C11 & C12 & C13 & C14 & C15 & C16 \\
\hline
Direct Distance & -0.3013 & 0.1697 & \textbf{0.9106} & -0.0966 & 0.1742 & -0.0920 & -0.0470 & -0.0146 & 0.0191 & -0.0007 & -0.0079 & 0.0009 & -0.0012 & 0.0171 & -0.0009 & 0.0005 \\
Road Distance & 0.2283 & -0.0926 & 0.1635 & 0.3187 & -0.0731 & -0.0344 & \textbf{0.7700} & -0.1680 & 0.3680 & 0.0025 & 0.0910 & 0.0113 & 0.0069 & 0.1987 & 0.0034 & 0.0055 \\
Num. Left Turns & \textbf{0.4457} & 0.0213 & 0.1532 & 0.0835 & -0.1044 & -0.2147 & -0.0823 & 0.0475 & -0.2461 & 0.0502 & 0.0436 & -0.2968 & 0.0148 & -0.0031 & -0.3152 & -0.6722 \\
Num. Right Turns & 0.3084 & -0.3444 & 0.1755 & 0.0794 & -0.1146 & -0.2337 & -0.0905 & 0.0292 & -0.2168 & -0.1006 & -0.0110 & 0.6109 & 0.0538 & -0.1288 & \textbf{0.4669} & -0.0667 \\
Num. Straight & 0.1994 & -0.0987 & -0.0950 & 0.0410 & 0.9622 & -0.0499 & 0.0662 & 0.0004 & -0.0797 & -0.0093 & -0.0120 & -0.0006 & 0.0081 & -0.0209 & 0.0028 & -0.0021 \\
Total Angle & 0.3007 & \textbf{0.8304} & -0.0566 & -0.0104 & 0.0210 & 0.0450 & 0.0081 & 0.0032 & 0.0601 & -0.3113 & -0.0037 & 0.3348 & 0.0019 & -0.0334 & -0.0194 & 0.0092 \\
Median Angle & \textbf{0.4437} & 0.0118 & 0.1536 & 0.0829 & -0.1033 & -0.2171 & -0.0836 & 0.0501 & -0.2489 & 0.0641 & 0.0356 & -0.2888 & 0.0202 & -0.0315 & -0.1391 & \textbf{0.7306} \\
Std Angle & 0.1117 & 0.2956 & -0.0191 & -0.0040 & 0.0119 & 0.0229 & -0.0105 & -0.0115 & 0.0170 & 0.4787 & -0.0104 & -0.3483 & -0.0090 & 0.0480 & \textbf{0.7329} & -0.0916 \\
Max Angle & 0.1514 & -0.0713 & 0.0231 & 0.1558 & 0.0425 & -0.0111 & -0.5247 & -0.0747 & \textbf{0.4051} & 0.0470 & 0.1824 & 0.0978 & 0.1140 & \textbf{0.6640} & -0.0328 & 0.0174 \\
Min Angle & 0.1366 & 0.0091 & 0.0453 & 0.1147 & 0.0176 & 0.0329 & -0.1839 & -0.1109 & \textbf{0.4188} & \textbf{0.5555} & 0.0046 & 0.2402 & -0.1757 & -0.5312 & -0.2520 & 0.0159 \\
Mean Angle & -0.0962 & 0.1302 & -0.0554 & -0.1244 & -0.0040 & -0.0033 & 0.2201 & 0.0609 & -0.4133 & \textbf{0.5867} & -0.0663 & 0.3854 & 0.1786 & 0.3766 & -0.2459 & 0.0294 \\
Median Pivot Off & -0.2875 & 0.1305 & -0.0771 & \textbf{0.6761} & 0.0135 & -0.1822 & -0.1126 & -0.3885 & -0.2939 & -0.0136 & -0.1116 & 0.0198 & -0.3706 & 0.0589 & 0.0060 & 0.0017 \\
Std Pivot Off & -0.2112 & 0.0925 & -0.0489 & \textbf{0.4256} & 0.0189 & -0.1034 & -0.0374 & 0.1325 & 0.0158 & 0.0081 & 0.1811 & -0.0445 & \textbf{0.7980} & -0.2478 & 0.0013 & -0.0070 \\
Max Pivot Off & 0.0913 & -0.0493 & 0.1334 & 0.1354 & 0.0043 & \textbf{0.6999} & -0.0222 & -0.0839 & -0.2883 & 0.0017 & \textbf{0.6017} & 0.0295 & -0.0778 & -0.0576 & 0.0177 & 0.0008 \\
Min Pivot Off & 0.0551 & -0.0323 & 0.1066 & 0.3956 & -0.0009 & 0.3609 & -0.0222 & \textbf{0.6912} & 0.0343 & 0.0069 & -0.4418 & 0.0097 & -0.1319 & 0.0746 & -0.0047 & -0.0006 \\
Mean Pivot Off & -0.1854 & 0.0752 & -0.1024 & 0.0329 & 0.0233 & -0.4151 & 0.0758 & \textbf{0.5394} & 0.0562 & 0.0419 & \textbf{0.5899} & 0.0507 & -0.3506 & 0.0126 & 0.0164 & 0.0010 \\
\hline
Importance & 30.96\% & 24.23\% & 13.80\% & 9.35\% & 5.74\% & 5.18\% & 3.38\% & 2.41\% & 1.81\% & 1.21\% & 0.70\% & 0.52\% & 0.34\% & 0.23\% & 0.10\% & 0.01\%\\
\hline
\end{tabular}
}
}
\end{table*}

\begin{table*}[t]
\centering
\caption{Results of the Principal Component Analysis (PCA) for Driver.AI. Values in boldface indicate the features that 
contribute the most to the main components (Cs) extracted by PCA.}
\vspace{-1mm}
\label{tab:pca:driveai}
\revise{
\resizebox{.95\textwidth}{!}{
    \begin{tabular}{c | rrrrrrrrrrrrrrrr}
\hline
Features  &  C1 & C2 & C3 & C4 & C5 & C6 & C7 & C8 & C9 & C10 & C11 & C12 & C13 & C14 & C15 & C16\\
\hline
Direct Distance & -0.1570 & 0.2939 & \textbf{0.8788} & -0.0933 & 0.3088 & -0.0926 & -0.0499 & -0.0153 & -0.0294 & 0.0084 & -0.0092 & 0.0064 & 0.0011 & 0.0173 & -0.0012 & -0.0003 \\
Road Distance & 0.1532 & -0.1709 & 0.1659 & 0.3213 & -0.0573 & -0.0128 & \textbf{0.7938} & -0.0402 & -0.3669 & 0.0512 & 0.0751 & 0.0055 & 0.0020 & 0.1961 & -0.0046 & 0.0027 \\
Num. Left Turns & 0.3838 & -0.1879 & 0.1669 & 0.1319 & -0.1174 & -0.2410 & -0.0898 & -0.0012 & 0.2592 & -0.0025 & 0.0384 & -0.2417 & 0.0110 & -0.0283 & -0.3206 & -0.6793 \\
Num. Right Turns & 0.0943 & -0.3614 & 0.1681 & 0.1273 & -0.1083 & -0.2190 & -0.0707 & -0.0137 & 0.1842 & -0.1474 & -0.0088 & \textbf{0.5874} & 0.0922 & -0.1185 & \textbf{0.5677} & -0.0751 \\
Num. Straight & 0.1414 & -0.2220 & -0.2185 & 0.1112 & \textbf{0.9269} & -0.0774 & 0.0262 & 0.0017 & 0.0606 & -0.0221 & -0.0078 & -0.0069 & 0.0142 & -0.0191 & 0.0005 & -0.0009 \\
Total Angle & \textbf{0.6742} & \textbf{0.5909} & -0.0893 & -0.0497 & 0.0323 & 0.0720 & 0.0043 & 0.0011 & -0.1245 & -0.2761 & -0.0155 & 0.2941 & 0.0234 & -0.0214 & -0.0070 & 0.0069 \\
Median Angle & 0.3858 & -0.1979 & 0.1719 & 0.1352 & -0.1190 & -0.2465 & -0.0922 & -0.0011 & 0.2679 & 0.0111 & 0.0378 & -0.2360 & 0.0154 & -0.0419 & -0.1552 & \textbf{0.7250} \\
Std Angle & 0.2334 & 0.2013 & -0.0285 & -0.0177 & 0.0139 & 0.0190 & 0.0112 & 0.0044 & 0.0838 & \textbf{0.4907} & 0.0118 & -0.4344 & -0.0506 & 0.0610 & \textbf{0.6736} & -0.0807 \\
Max Angle & 0.0801 & -0.1131 & 0.0146 & 0.1994 & -0.0055 & -0.0443 & -0.5054 & -0.0193 & -0.4072 & 0.1097 & 0.1743 & 0.0810 & 0.1024 & \textbf{0.6764} & -0.0111 & 0.0062 \\
Min Angle & 0.1119 & -0.0437 & 0.0299 & 0.1272 & 0.0011 & 0.0117 & -0.1629 & -0.0436 & -0.3222 & \textbf{0.6446} & 0.0343 & 0.3028 & -0.1618 & -0.5037 & -0.2181 & 0.0148 \\
Mean Angle & -0.0180 & 0.1361 & -0.0565 & -0.1477 & 0.0191 & 0.0074 & 0.2188 & 0.0212 & \textbf{0.5247} & 0.4732 & -0.0530 & 0.3907 & 0.1923 & 0.4074 & -0.2220 & 0.0220 \\
Median Pivot Off & -0.2307 & 0.3422 & -0.1052 & \textbf{0.6554} & -0.0080 & -0.1726 & -0.0473 & -0.4003 & 0.2179 & -0.0509 & -0.0995 & 0.0373 & -0.3643 & 0.0513 & 0.0027 & 0.0011 \\
Std Pivot Off & -0.1630 & 0.2366 & -0.0680 & 0.3908 & -0.0040 & -0.0959 & -0.0200 & 0.1901 & -0.0124 & 0.0239 & 0.1577 & -0.0874 & \textbf{0.7919} & -0.2322 & 0.0001 & -0.0037 \\
Max Pivot Off & 0.0615 & -0.0875 & 0.1406 & 0.2016 & 0.0269 & \textbf{0.7212} & -0.0486 & -0.0572 & 0.2615 & -0.0544 & \textbf{0.5681} & 0.0246 & -0.0673 & -0.0476 & 0.0085 & -0.0011 \\
Min Pivot Off & 0.0280 & -0.0266 & 0.0891 & 0.3459 & -0.0095 & 0.3006 & -0.0711 & \textbf{0.6949} & 0.0570 & -0.0001 & -0.5096 & 0.0150 & -0.1592 & 0.0626 & -0.0064 & 0.0009 \\
Mean Pivot Off & -0.1191 & 0.1603 & -0.1018 & -0.0265 & 0.0076 & -0.4015 & 0.0712 & \textbf{0.5591} & 0.0349 & -0.0167 & \textbf{0.5824} & 0.0687 & -0.3527 & 0.0128 & 0.0091 & 0.0003 \\
\hline
Importance & 29.97\% & 25.43\% & 12.33\% & 9.34\% & 7.09\% & 5.25\% & 3.60\% & 2.32\% & 1.67\% & 1.18\% & 0.77\% & 0.49\% & 0.38\% & 0.24\% & 0.10\% & 0.01\%\\
\hline
\end{tabular}
}
}
\end{table*}

\section{Results}
\label{sec:results}
This section reports, for each research question, the obtained results and main findings.

\subsection{
RQ$_1$: SDC Features Analysis}
\label{sec:results:rq1}



Tables \ref{tab:pca:beamng1}, \ref{tab:pca:beamng15}, and \ref{tab:pca:driveai} show the results of the PCA for datasets \texttt{BeamNG.AI.AF1}, \texttt{BeamNG.AI.AF1.5}, and \texttt{Driver.AI}, respectively. 
 As we can observe, for each dataset, PCA identifies 16 (independent) principal components, whose relative importance is reported on the last (bottom) row of the corresponding Table. As these rows indicate, the importance of components in all of the tables (\ie datasets) are similar: the first component (C1) covers about 30\% of the variance in the data (importance), followed by the second components (C2) with about 25\%, and so on. Moreover, in all of the datasets, the last six principal components are negligible as they contribute to less than 1\%  of the total variance.

Looking at the scores achieved for the different features, we can observe that they contribute to different (orthogonal) latent components. Hence, the features capture different characteristics of the road segments in the test scenarios. Individual features exclusively capture certain components. For example, in Table \ref{tab:pca:beamng1}, C2 (which corresponds to 26\% of the proportion) is fully captured by the feature F6 (i.e., number of turns) with a score greater than 79\%. Similar observations can be made for other components: C3 (14\% of importance) is captured by F1 (direct end-to-end distance) with an 87\% score; C5 (6\% of importance) is exclusively related to F5 (number of straight segments) with 96\% score; and so on. Similar results can also be observed in Tables \ref{tab:pca:beamng15} and \ref{tab:pca:driveai}.

Closely looking at C1, C9, and C10, in Table \ref{tab:pca:beamng1}, (or C1, C4 in Table \ref{tab:pca:beamng15} and C8 and C10 in Table \ref{tab:pca:driveai}) we can observe that there are at least two features that equally contribute to them. In other words, some road features show some degree of collinearity. Finally, Features F3 (number of left turns) and F7 (median angle of turns in the road) both contributed about 40\%  to the first components (C1), which is the most important component according to PCA.

Therefore, we can conclude that the designed road features show some level of multi-collinearity, which is limited to a few features and for a few latent components. Hence, we use PCA for dimensionality reduction and feature extraction as described in Section~\ref{sec:pca}. In particular, we select the top $h=10$ principal components as they correspond to (cumulatively) 98\% of the original data variance. According to the PCA Tables, the last six components are negligible as together account for less than 2\% of the data variance in all of the datasets. 

Given the results above, we used the lower-dimensional $M'$ matrix produced by the PCA with $h=10$ to compute the Euclidean distances and the fitness function used by \approach and greedy-based test prioritization in RQ$_2$ and RQ$_3$. In particular, we use the new set of (non-collinear) features obtained with Equation~\ref{eq:pca}.

\finding{1}{The designed road features show some level of multi-collinearity. The first ten principal components produced by PCA allowed the identification of the ten meta-features,  representing 98\% of the original datasets' variance, to consider for experimenting with prioritization strategies (i.e., RQ$_2$).}

\begin{figure}[t]
    \centering
    \revise{
    \frame{
	\includegraphics[width=0.85\linewidth]{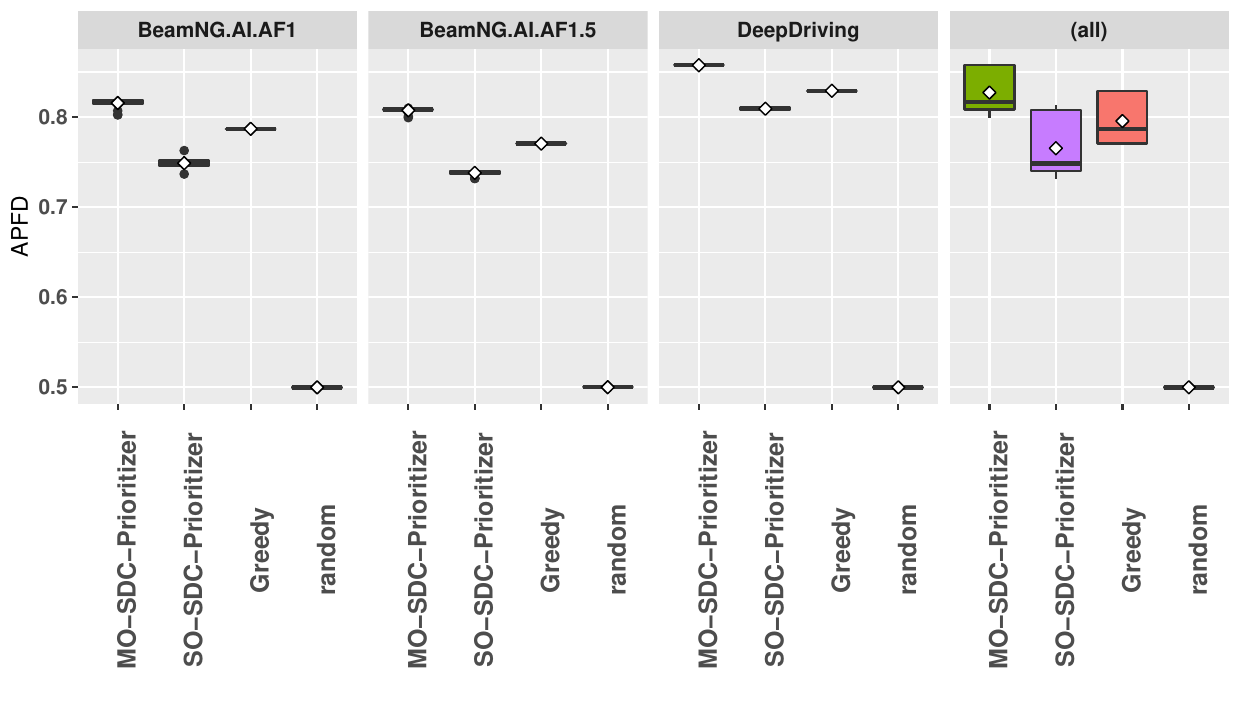}
	}
	}
	\caption{APFD$_c$ achieved by \revise{\sapproach, \mapproach, and greedy approach with ten features and random test prioritization}. The diamond ($\diamond$) denotes the arithmetic mean, and the bold line (---) is the median.}
  \label{fig:APFD}
\end{figure}

\subsection{
RQ$_2$: Cost-effectiveness of \approach Compared to Baseline Approaches
}
\label{sec:results:rq2}

\begin{figure*}[t]
\subfloat[BeamNG.AI.AF1]{
\revise{    
\frame{
\includegraphics[width=0.32\textwidth]{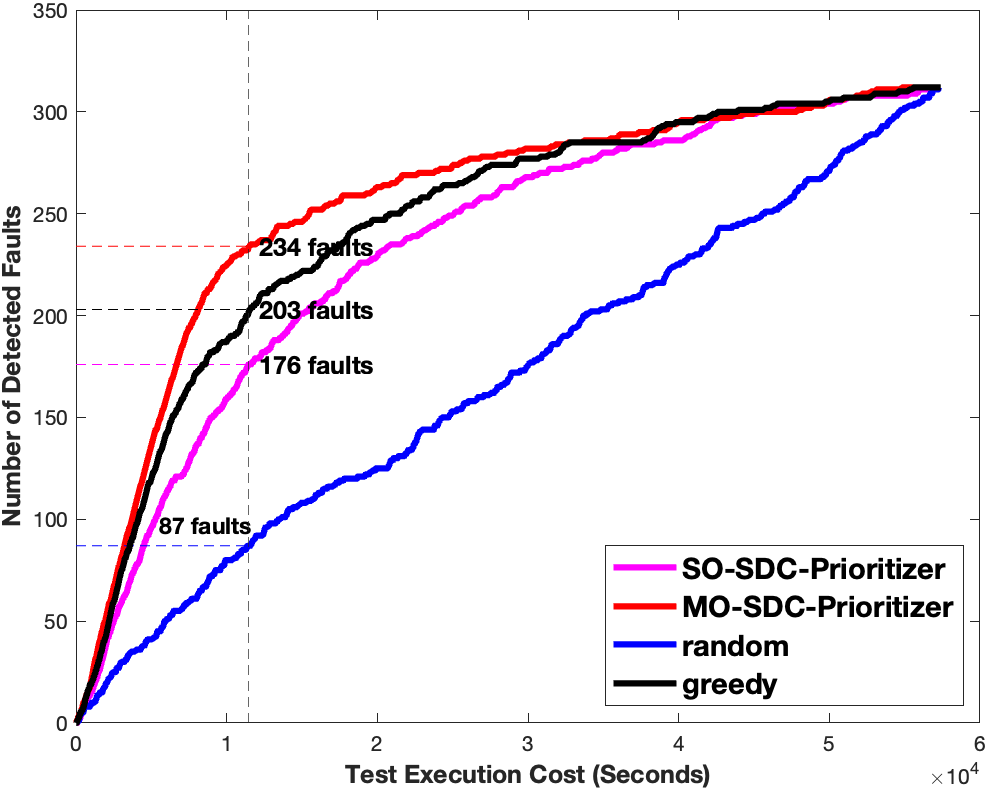}
}}
\label{fig:samples:AI1}
}
\subfloat[BeamNG.AI.AF1.5]{
\revise{    
\frame{
\includegraphics[width=0.32\textwidth]{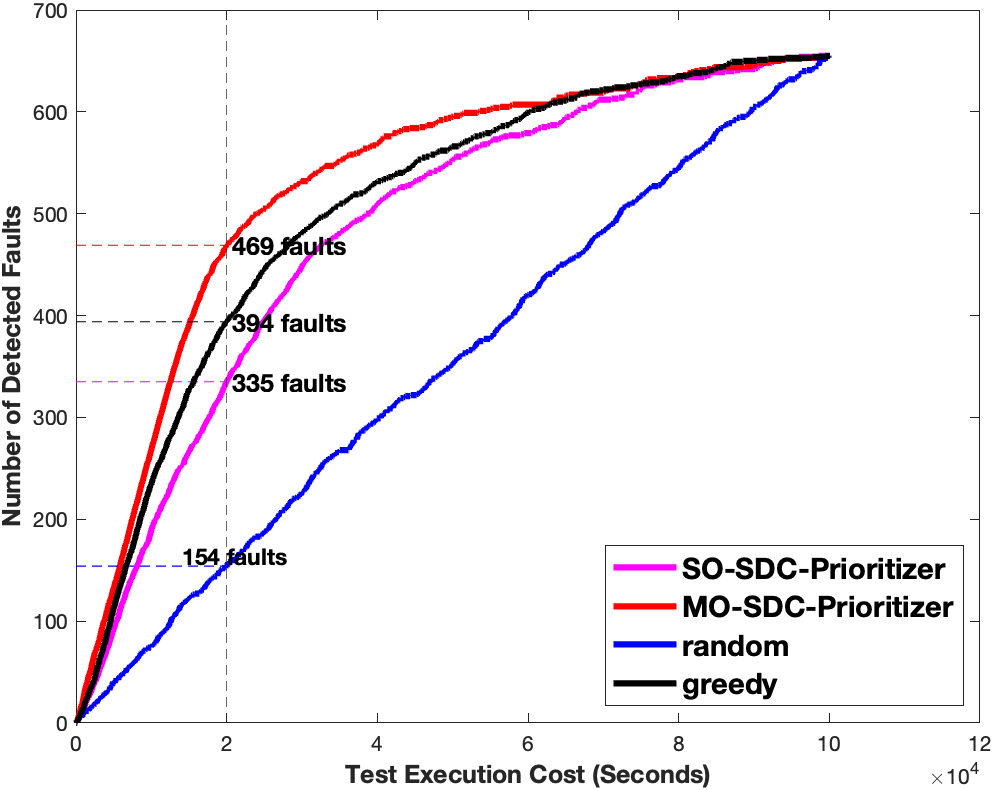}
}}
\label{fig:samples:AI15}
}
\subfloat[Driver.AI]{
\revise{    
\frame{
\includegraphics[width=.32\textwidth]{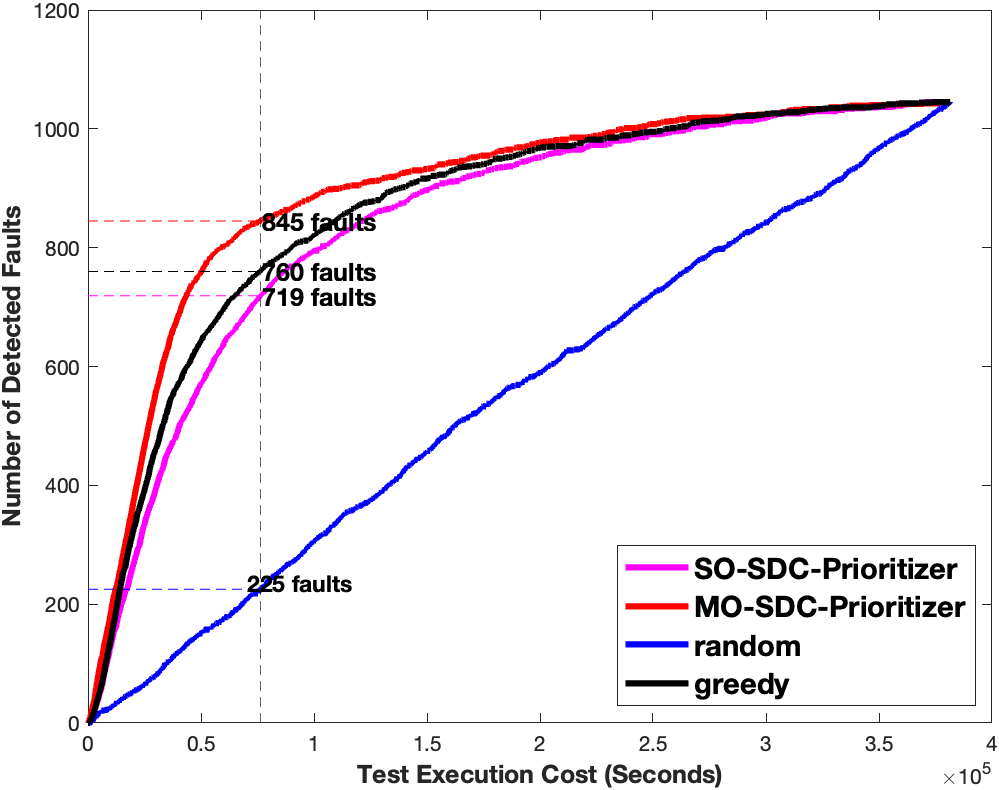}
}}
\label{fig:samples:DeepDriving}
}
    \vspace{-2mm}
\caption{Cost-effectiveness curves produced by the different TPC methods. Each curve depicts the cumulative number of detected faults the cumulative test execution costs yielded by the test case prioritizations. }
\label{fig:samples}
\end{figure*}

This section compares \revise{\sapproach, \mapproach,} random and greedy-based test prioritizations in terms of \apfdc. 
\revise{
For both  \approach and the greedy-based approach, we use the first 10 principal components produced by PCA (detailed in Section 5.1). This allows us to perform an unbiased evaluation.
We do not use these features nor the PCA for random search since (unlike \approach and greedy) it does not require features to measure the distance between two tests.
}
Figure \ref{fig:APFD} depicts the \apfdc values achieved by \revise{\sapproach, \mapproach}, greedy-based, and random test prioritization approaches. As we can see in this figure, \revise{the best performing test prioritization in all of the datasets is \mapproach. In each dataset, the minimum \apfdc achieved by \mapproach is higher than the maximum \apfdc achieved by other test prioritization configurations.
In all three datasets, the minimum \apfdc achieved by \mapproach is at least 2\%, 4\%, 30\% is higher than the highest \apfdc produced by greedy, \sapproach, and random test prioritization, respectively. 
On average, \mapproach reaches about 3\%, 6\%, and 25.5\% higher \apfdc than Greedy, \sapproach, and random test prioritization, respectively. \revise{The second-best test prioritization technique is the greedy search (achieving an average \apfdc of  79.5\%), followed by \sapproach (with an average \apfdc of 76.5\%) and random test prioritization (with an average \apfdc of 49.9\%)}.
}

Moreover, as reported in Table \ref{tab:ga-vs-baseline}, \revise{\mapproach} significantly (p-values$ < 1.0 e-10$) outperforms (as all $\hat{A}_{12}$ values are all higher than 0.5) both random and greedy test prioritization  in terms of \apfdc score. The magnitude of the difference (effect size) is \textit{large} in all datasets. \revise{Same as \mapproach, \sapproach significantly outperforms random test prioritization. However, this test prioritization technique achieves significantly lower \apfdc values in comparison with greedy-based test prioritization in all datasets.}
\revise{Similar to the pairwise comparison of \approach variants with baselines, \mapproach significantly achieves higher \apfdc than \sapproach in all datasets (p-values$ < 1.0 e-10$, $\hat{A}_{12} = 1$, and large magnitude of effect sizes).}

To provide more insights into these results, we graphically compare the cumulative number of faults detected by the different approaches when running the test cases incrementally according to the test prioritizations they produced. 
For each dataset, we took a more detailed look at the permutations generated by \revise{ each \approach variant} that achieve an \apfdc value equal to the median of the \apfdc values delivered by all applications of \revise{that \approach variant} on a specific dataset. Specifically, \revise{for each of the \sapproach and \mapproach}, we sampled three permutations generated by \revise{these techniques} for each of the datasets. For each dataset, we compare the sampled permutation\revise{s} against the best output of random (\ie the permutation generated by random that gains the best \apfdc) and greedy strategies. For this comparison, we analyze the rate of fault occurrences during the execution of tests, according to the generated permutations.

\begin{table}[t]
\caption{Comparison of APFD$_c$ score achieved by \revise{\sapproach and \mapproach against} the baselines, for each of the datasets used in this study. $p$-values for Wilcoxon tests, Vargha Delaney's estimates ($\hat{A}_{12}$), and magnitudes are reported.}
\label{tab:ga-vs-baseline}
\small
\revise{
\begin{tabular}{ l l | c c c | c c c} \\
\hline 
 \textbf{GA Config.} & \textbf{Dataset} & \multicolumn{3}{c|}{\textbf{Vs. Random}} & \multicolumn{3}{c}{\textbf{Vs. Greedy}} \\
  &   & $\hat{A}_{12}$ & p & Magnitude & $\hat{A}_{12}$ & p & Magnitude \\ 
\hline 
 & BeamNG.AI.AF1  & 1.0 & 3.016e-11 & large & 1.0 & 1.21e-12 & large \\ 
\textbf{MO-SDC-Prioritizer}& BeamNG.AI.AF1.5  & 1.0 & 3.016e-11 & large & 1.0 & 1.211e-12 & large \\ 
&DeepDriving & 1.0 & 2.113e-11 & large & 1.0 & 7.602e-13 & large \\ 
\hline
 & BeamNG.AI.AF1 & 1.0 & 3.018e-11 & large & 0.0 & 1.211e-12 & large \\ 
\textbf{SO-SDC-Prioritizer} & BeamNG.AI.AF1.5 & 1.0 & 3.018e-11 & large & 0.0 & 1.212e-12 & large \\ 
 & DeepDriving & 1.0 & 3.018e-11 & large & 0.0 & 1.211e-12 & large \\ 
\hline 
\end{tabular}
}
\end{table} 

Figure \ref{fig:samples} depicts this comparison for each dataset. As we can see from the figure, in all of the benchmarks, running the tests using the test case orders generated by \revise{\mapproach} leads to a higher rate of fault occurrence in a shorter time. 
As a concrete example, in this figure, we highlighted the number of faults that occurred with the first 20\% of the test execution.
In the dataset \textsc{BeamNG.AI.AF1} (Figure \ref{fig:samples:AI1}), the permutation generated by \revise{\mapproach} leads to the detection of \revise{234} faults in the first 20\% of test execution time. \revise{This value reduces for greedy (203), \sapproach (176), and random (87) test prioritization approaches.}
Similarly, in the second dataset (Figure \ref{fig:samples:AI15}), \revise{\mapproach} generates a permutation, which is able to detect \revise{469} faults in the first 20\% of the test execution. Also, in this case, this number is lower for the other approaches: \revise{394, 335, and 154 faults detected by the greedy, \sapproach, and random approaches, respectively.}
The same trend is observed in the dataset of \textsc{Driver.AI} (Figure \ref{fig:samples:DeepDriving}), in which, the sampled permutation from \revise{\mapproach} can detect \revise{845} faults, i.e., \revise{+85, +126, and +620 more faults compared to greedy, \sapproach, and random algorithms}, respectively.


 \finding{2}{\revise{\mapproach} increases the \apfdc score on average compared with random and greedy approaches. The improvement achieved by \approach, in terms of fault detection rate, is statistically significant. \revise{Unlike \mapproach, which is the best performing test prioritization technique in terms of fault detection capability, \sapproach only achieves higher \apfdc than random approach. This observation stems from the lack of exploration ability in this single-objective meta-heuristic, which drives the search process to trap local optima.}}

\begin{figure*}[t]
    \centering
    \revise{    
    \frame{
	\includegraphics[width=0.85\linewidth]{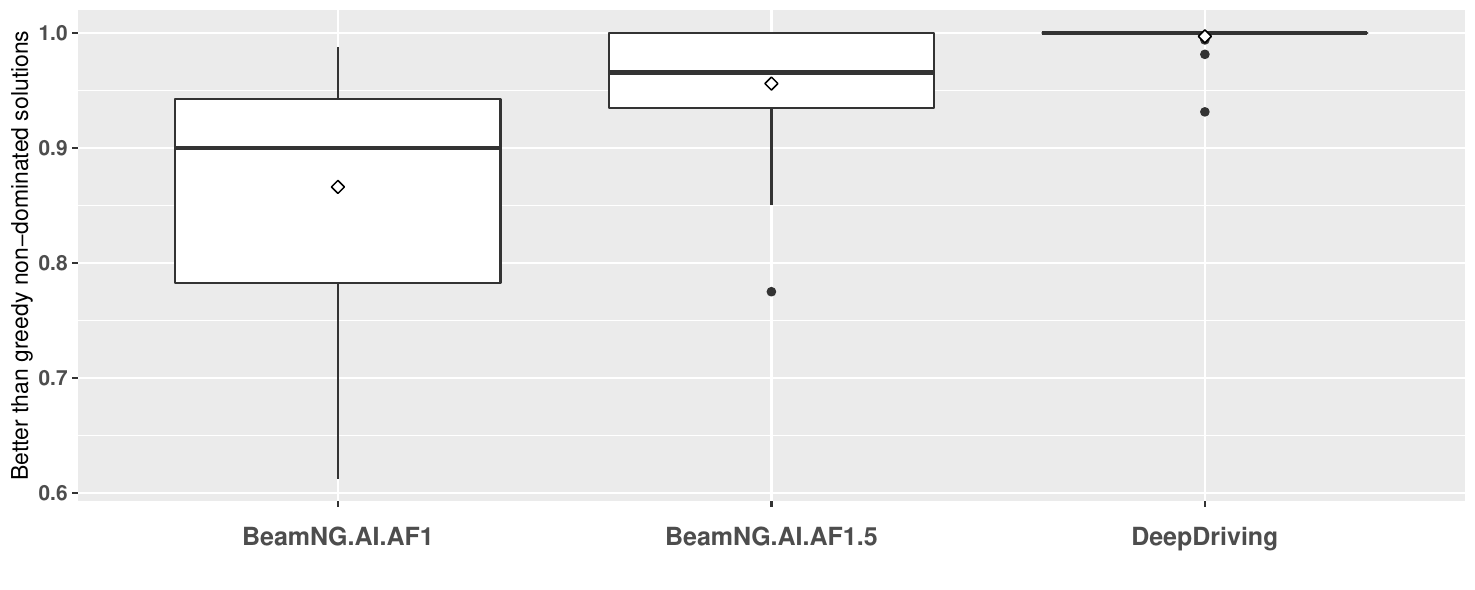}
	}
	}
	\caption{\revise{Percentage of non-dominated solutions generated by \mapproach that achieve a higher \apfdc compared to greedy-based test prioritization.}}
 \label{fig:better_rate_plo}
\end{figure*}

\begin{figure*}[t]

    \centering
    \revise{    
    \frame{
	\includegraphics[width=0.85\linewidth]{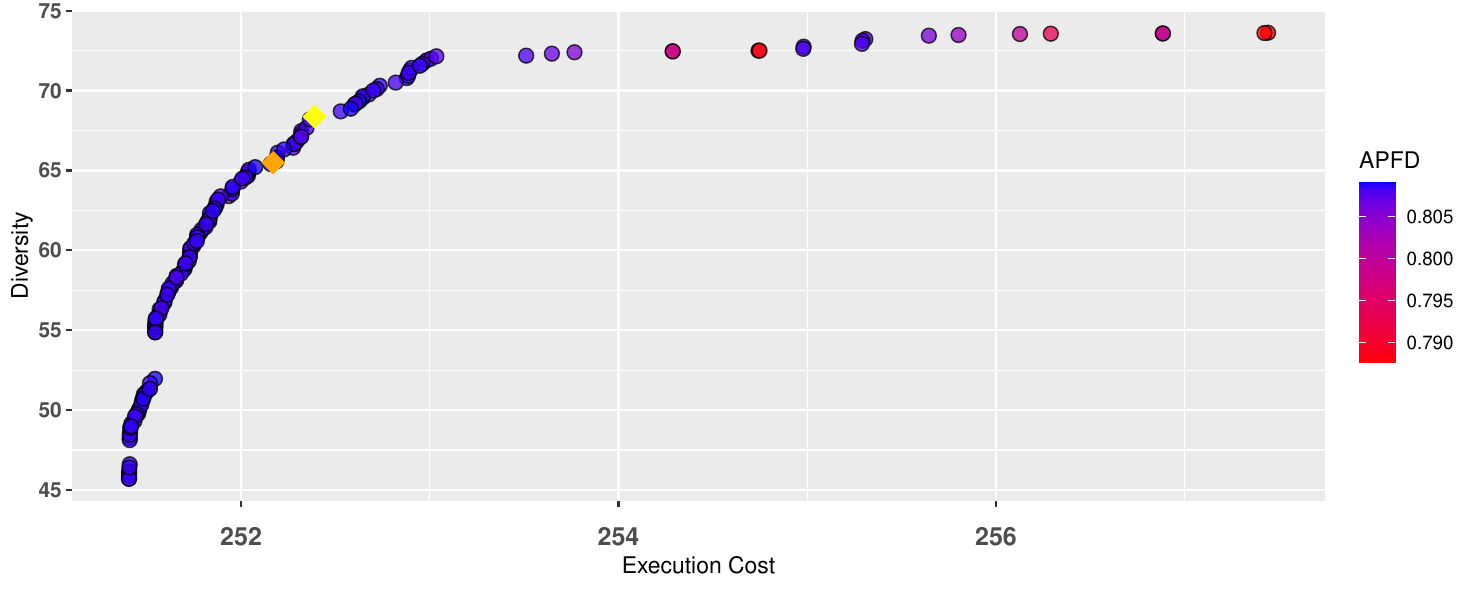}
	}}
	\caption{\revise{A sample of Pareto front generated by \mapproach in BeamNG.AI.AF1.5 dataset. Each circle point represents one of the non-dominated solutions in the Pareto front. The blue points are the solutions with an \apfdc score larger than the one produced by the greedy algorithm. The orange and yellow diamond points indicate the solution with the highest APFD and the closest solution to the utopia point, respectively.}}
 \label{fig:samples:pareto}
\end{figure*}

\subsubsection{\revise{Pareto fronts in \mapproach}}
\label{sec:paretofront}

\revise{
As explained in Section \ref{sec:approach}, same as any other multi-objective approaches, \mapproach returns a set of non-dominated solutions in output. To answer RQ$_2$, we selected the closest non-dominated solution to the utopia point (explained in Section \ref{sec:approach:select:solution}). Results presented by this section indicated that this solution has higher \apfdc compared to the test execution orders generated by other techniques. However, we perform a more in-depth analysis to understand whether other non-dominated solutions could be selected from the Pareto front. To this aim, we compare  the Pareto fronts (\ie non-dominated test orders) generated by each \mapproach's run with the \apfdc achieved by the second-best technique (\ie greedy-based test prioritization) in terms of fault detection capability. Figure \ref{fig:better_rate_plo} presents the percentage of non-dominated solutions generated by \mapproach that achieves a higher \apfdc compared to the Greedy approach. On average, about 94\% of non-dominated solutions generated by \mapproach can detect more unsafe tests than Greedy and in shorter times (\ie they have higher \apfdc).
Even in the worst scenario (17th execution of \mapproach on BeamNG.AI.AF1 dataset), more than 61\% of generated solutions in the final Pareto front produced by \mapproach has higher \apfdc compared to Greedy. The highest performance of \mapproach can be observed when this test prioritization technique is utilized to prioritize tests for the DeerDriving dataset in which, on average, 99.7\% of solutions have higher \apfdc than the ones generated by Greedy test prioritization.
}

\revise{To better understand the impacting factors that lead the generated non-dominated solutions to achieve a high \apfdc, we manually analyzed the \apfdc values of Pareto fronts generated by \mapproach in each dataset. In all of the cases, we observed the same trend as the sample, presented in Figure \ref{fig:samples:pareto}. This figure is a two-dimensional vector in which each dimension indicates one of the \mapproach's search objectives (diversity and execution cost). As we can see, all solutions with the lowest \apfdc (red points in the Pareto front) are the extreme points with the maximum diversity and maximum test execution costs. In addition, the solution with the highest \apfdc (the orange diamond point) is not in the extreme parts of the Pareto front (\ie it has a good balance between the diversity and execution cost). As we can observe, the knee point selected by \mapproach is among the middle points in the front with the largest \apfdc. Besides, it is very close to the best point (in terms of \apfdc) within the Pareto front. This observation empirically supports the technique we used for selecting the final test order (the yellow diamond point).}

\finding{3}{\revise{On average, the majority (94\%) of the solutions generated by \mapproach has higher \apfdc than Greedy (the second-best test prioritization technique for detecting faults in a shorter time). By taking a deeper look at non-dominated solutions generated by \mapproach, we can see that the few solutions with lower \apfdc are at the extremes of the Pareto front. Moreover, the solutions with the highest \apfdc values are the ones that have a balance between tests diversity and test execution cost.}}

\subsection{
RQ$_3$: Overhead of \approach
}

\begin{figure*}[t]
    \centering
    \revise{
    \frame{
	\includegraphics[width=0.7\linewidth]{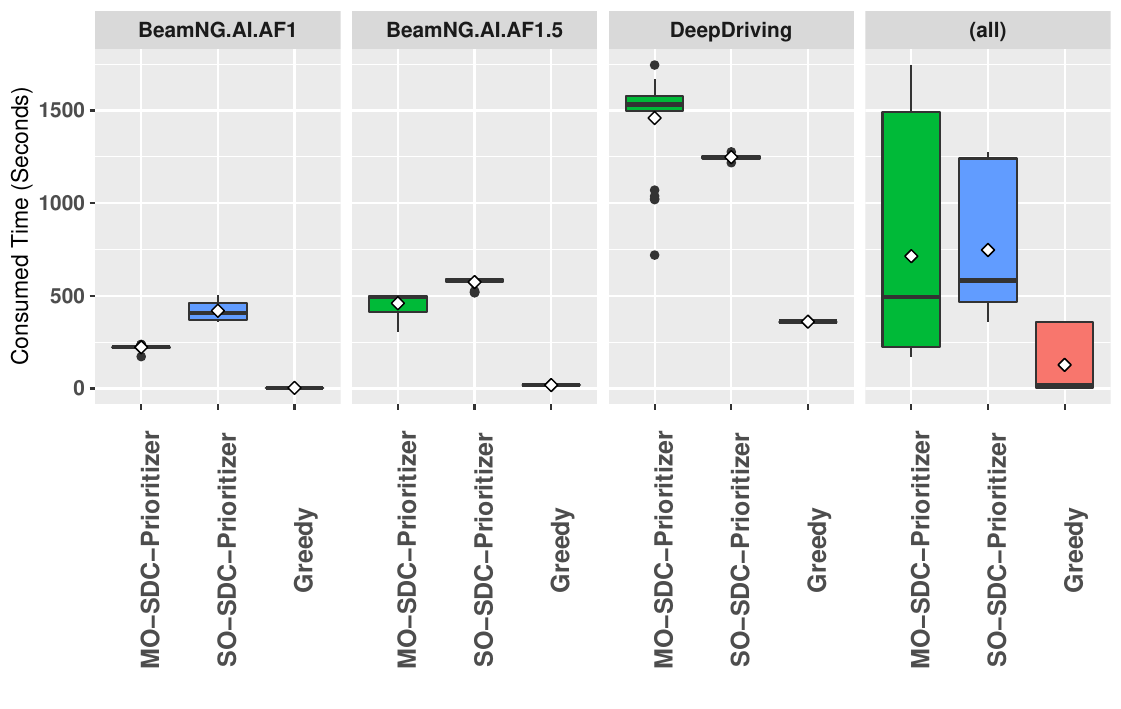}
	}
	}
	\caption{Running Time of the different TCP approaches}
  \label{fig:consumed_time}
\end{figure*}

Figure \ref{fig:consumed_time} illustrates the distribution of the time consumed by \revise{\sapproach, \mapproach, } and greedy test prioritization. \revise{As this figure shows, on average, \revise{\sapproach and \mapproach} require about \revise{12.5} and \revise{11.5} minutes to finish the search process with 4,000 generations, \revise{respectively}.} Practically, this amount of time is negligible if we consider the total 16 to 106 hours needed to run the entire set of tests, and that  \revise{ both variants of \approach do} not negatively impact the performance (e.g., on fault detection) of testing practices. In fact, the overall overhead accounts for \revise{0.38\%} (for Driver.AI) and a maximum of \revise{0.45\%} (for BeamNG.AI.AF1.5) of the cost needed to run the entire test suites.

\finding{4}{The overhead introduced by \revise{each \approach variants} is less than \revise{13} minutes and is imperceptible for an SDC simulation pipeline used by developers to test the SDCs behavior in critical scenarios.}
  
Figure \ref{fig:consumed_time} shows that (right side of the Figure) the average time required by the greedy approach is about \revise{five times} shorter than what \revise{\sapproach or \mapproach} needs. Even though \revise{\mapproach} is slower than greedy (\ie it needs about \revise{10} minutes more time), it performs better in terms of \apfdc score (as shown by Section \ref{sec:results:rq2}).

\finding{5}{On average, \mapproach needs about \revise{10} minutes more than the greedy test prioritization. However, this negligible extra overhead significantly increases the \apfdc values achieved by the subsequently generated test prioritization.}
   
\revise{Finally, it is worth mentioning that \approach techniques include two main parts:} (i) pairwise comparison of distances between every two tests (using Euclidean distance), and (ii) running the genetic algorithm. The former is a one-time task (\ie by one execution, we can run the genetic algorithm multiple times) with the time complexity of $O(n^2)$, where $n$ is the number of tests. Since the latter part uses the values calculated in pairwise distance calculation for fitness function evaluation, the complexity of this task is $O(n)$ (this complexity is due to the search for the most diverse test). Also, the time complexities of mutation and crossover operators are $O(n)$. Hence, \approach has $O(n^2)$ one-time cost (for calculating the distances) and $O(n \times m)$ for the whole search process, where $n$ is the number of tests, and $m$ is the number of fitness evaluations. According to this information, we can confirm that \approach scales for a large-size test set. 
\revise{Similarly, the test suites used in our study are much larger than the other ones reported in prior studies on regression testing \cite{Rothermel:icsm1999}}. Our largest test suite (Driver.AI) contains 5,630 tests. On average, \approach \revise{approaches} performed the test prioritization for this test suite in less than \revise{25} minutes.

\section{Threats to Validity}
\label{sec:threats}

Threats to \emph{construct validity} concern the relationship between theory and observation. In this case, threats can be mainly due to the imprecision in simulation realism as well as the automated classification of safe and unsafe scenarios. We mitigated both threats by leveraging BeamNG (used in this year`s SBST tool competition \cite{SBST2021}) as a soft-body simulation environment (which ensures a high simulation accuracy in safety-critical scenarios) and \framework \cite{Birchler:2022} (which integrates also AsFault) as a technological reference solution to generate and execute test cases, as detailed in Section \ref{sec:study}. \revise{Furthermore, to address the potential threat to have high variability in execution time of the executed tests, we selected a sample of 50 test cases (using a stratified random sampling, equal distribution of safe and unsafe tests)  and executed them 10 times each. As mentioned in Section \ref{sec:cos-effectivenes}, the standard deviation of the execution time is negligible.}

Threats to \emph{internal validity} may concern, as for previous work~\cite{GambiMF19}, the relationships between the technologies used to generate the scenarios and the realism of simulation results. Specifically, we did not recreate all the elements that can be found on real roads (e.g., weather conditions, light conditions, etc.). However, to increase our internal validity, we focused on the usage of both BeamNG.AI and Driver.AI as test subjects. This allows us to assess the cost-effectiveness of our approach by experimenting with different driving styles and driving risk levels. Both BeamNG.AI and Driver.AI leverage a good knowledge of the roads, which means that they do not suffer from limitations of vision-based lane-keeping systems. 
\revise{However, since with BeamNG.AI it is possible to adjust the driving risk level, a higher amount of unsafe test scenarios can be observed. Hence, an AI implemented in physical SDC might be much more conservative in its driving style, which is something we plan to investigate for future work.}

Finally, threats to \emph{external validity} concern the generalization of our findings. The number of experimented test case scenarios in our study is larger than in previous studies \cite{GambiMF19} and we experimented with different AI engines. However, our results could not generalize with the universe of general open-source CPS simulation environments used in other domains. Therefore,  further studies considering more SDC data, other CPS domains, and different safety requirements are expected.
To minimize potential external validity in our evaluation setting, we followed the guidelines by Arcuri {\em et al.} \cite{ArcuriB14}: we compared the results of \approach with randomized test generation algorithms (the baseline approaches described in Section \ref{sec:study}) presented and repeated the experiment 30 times. Finally, we applied sound non-parametric statistical tests and statistics to analyze the achieved results.

\section{Conclusions \& Future Work}
\label{sec:conclusions}
Regression testing for self-driving cars (SDCs) is particularly expensive due to the cost of running many test driving scenarios  (test cases) that interact with simulation engines. To improve the cost-effectiveness of regression testing, we introduced two black-box test case prioritization approaches, called \revise{\sapproach and \mapproach}. These approaches rely on a set of static road features and are suitably designed for SDCs. These features can be extracted from the driving scenarios prior to running the tests. 
\revise{Both of these techniques utilize} genetic algorithms (GAs) to prioritize the test cases based on their distances (diversity) computed using the proposed road features and \revise{test execution costs}. \revise{\sapproach performs a single-objective optimization to fulfill this task (\ie both test diversity and execution costs are included in a single fitness function), while \mapproach leverages one of the common multi-objective genetic algorithms (NSGA-II) to prioritize tests according to two search objectives (one for differences of tests and the other one for test execution costs).}

  
We empirically investigated the performances of \revise{\sapproach and \mapproach} and compared it with two baselines: random search and greedy algorithms. 
Finally, we assessed whether \revise{these proposed techniques do} not introduce a too large computational overhead to the regression testing process.
Our results show that \revise{\mapproach} is more cost-effective than the baseline approaches. Specifically, the single solution provided by \mapproach dominates the solutions provided by \revise{\sapproach and} the baselines in terms of test execution time and fault detection capability.
Moreover, \revise{both \approach techniques} successfully prioritize the test cases independently of which AI engine is used (i.e., Driver.AI and BeamNG.AI) or different risk levels (i.e., different driving styles). Interestingly, looking at the running time, we can observe that the overhead required by \revise{\sapproach and \mapproach} in prioritizing the test scenarios is negligible with regards to the overall test execution cost.

We plan to replicate our study on further SDC AIs and additional SDC features as future work. Moreover, we plan to perform new empirical studies on further CPS domains to investigate additional safety criteria \revise{concerning new types of faults different from those investigated in this work. Specifically, important for this is to investigate approaches that are more human-oriented or are able to integrate humans into-the-loop~\cite{washingtonpost:2019,PanichellaSGVCG15,SorboPASVCG16,Grano:2017}. Moreover, we want to investigate different meta-heuristics in addition to the GA used in this paper. \revise{Complementary, we aim to investigate different distance functions to measure the diversity of the test cases (e.g., graph-based distances over feature-vector-based distances). Finally, we plan to integrate the proposed solution based on the experimented simulation environments to pririotize devise signals into industrial context such as AICAS context\footnote{https://www.aicas.com/wp/}, involved in the COSMOS H2020 project\footnote{https://www.cosmos-devops.org/}.
}
}

\subsection*{Acknowledgements \& Credit Author Statement}
We gratefully acknowledge the Horizon 2020 (EU Commission) support for the project \textit{COSMOS} (DevOps for Complex Cyber-physical Systems), Project No. 957254-COSMOS. 

\balance
\bibliographystyle{ACM-Reference-Format}
\bibliography{references}

\end{document}

%% file: graphs/dominance.tex
\begin{tikzpicture}
    \tikzset{
        hatch distance/.store in=\hatchdistance,
        hatch distance=10pt,
        hatch thickness/.store in=\hatchthickness,
        hatch thickness=2pt
    }

    \makeatletter
    \pgfdeclarepatternformonly[\hatchdistance,\hatchthickness]{flexible hatch}
    {\pgfqpoint{0pt}{0pt}}
    {\pgfqpoint{\hatchdistance}{\hatchdistance}}
    {\pgfpoint{\hatchdistance-1pt}{\hatchdistance-1pt}}%
    {
        \pgfsetcolor{\tikz@pattern@color}
        \pgfsetlinewidth{\hatchthickness}
        \pgfpathmoveto{\pgfqpoint{0pt}{0pt}}
        \pgfpathlineto{\pgfqpoint{\hatchdistance}{\hatchdistance}}
        \pgfusepath{stroke}
    }
        \begin{axis}[
            axis x line = bottom,
            axis y line = left,
            xmin=0, xmax=125,
            ymin=0, ymax=100,
            ylabel= \textbf{Diversity ($f_1$)},
            xlabel=\textbf{Cost ($f_2$)},
              width=0.7\linewidth,height=0.4\linewidth,
            ]
        \addplot[fill=lightgray, opacity=0.3] coordinates
        {(0,40) (70,40) (70,100) (0, 100)}
        -| (current plot begin)
        \closedcycle;
        \addplot[color=gray, fill = blue, opacity=0.2] coordinates
        {(50,60) (150,60)}
        \closedcycle;
        
        \addplot[only marks] coordinates
        {(25,20)}
        \closedcycle;
        \addplot[only marks, nodes near coords=A] coordinates
        {(25,20)}
        \closedcycle;
        
        \addplot[only marks] coordinates
        {(50,60)}
        \closedcycle;
        \addplot[only marks, nodes near coords=B] coordinates
        {(50,60)}
        \closedcycle;
        
        \addplot[only marks] coordinates
        {(75,65)}
        \closedcycle;
        \addplot[only marks, nodes near coords=C] coordinates
        {(75,65)}
        \closedcycle;
        
        \addplot[only marks, mark=square*, mark options={fill=white},scale=2] coordinates
        {(100,40)}
        \closedcycle;
        \addplot[only marks, nodes near coords=\textbf{E}, every node near coord/.style={anchor=180}] coordinates
        {(100,40)}
        \closedcycle;
        
        \addplot[only marks, mark=square*, mark options={fill=white},scale=2] coordinates
        {(70,40)}
        \closedcycle;
        \addplot[only marks, nodes near coords=\textbf{D}, every node near coord/.style={anchor=180}] coordinates
        {(70,40)}
        \closedcycle;
        \end{axis}
    \end{tikzpicture}

%% file: graphs/knee-point.tex
\pgfplotsset{compat=newest}

\begin{tikzpicture}

\begin{axis}[
       xlabel={Cost},
     ylabel={Diversity},
    xmin=0,  xmax = 1,
    xtick = {0,0.1,...,1.1},
    ymin=0, ymax =1,
    ytick = {0,0.2,...,1},
    axis lines = left,
    width=0.6\linewidth,height=0.4\linewidth,
    colormap={my colormap}{
                color=(ACMBlue)
                color=(ACMRed)
            },
]
\addplot+ [
        scatter,
        ACMBlue,
        scatter src=explicit,
]
coordinates {
	(0.2, 0.10) [1]
	(0.23, 0.25) [1]
	(0.28, 0.35) [1]
	(0.35, 0.48) [1]
	(0.45, 0.65) [4]
	(0.6, 0.75) [1]
	(0.7, 0.78) [1]
	(0.9, 0.85) [1]
};
\draw [ dashed] 
        (axis cs: 0.2, 0) -- (axis cs: 0.2,0.85)
        node[pos=0.5, above] {};
\draw [ dashed] 
        (axis cs: 0.2,0.85) -- (axis cs: 1.00,0.85)
        node[pos=0.5, above] {};
\addplot [black, mark = *, nodes near coords=\small Utopia, every node near
coord/.style={anchor=270}] coordinates {( 0.2, 0.85)};
\node (source) at (axis cs: 0.2, 0.85){};
\node (destination) at (axis cs:0.45, 0.65){};
\draw [dashed][->](source)--(destination);
\node at (axis cs:0.45, 0.65) [anchor=170] {\small Knee Point};
\node at (axis cs:0.20, 0.10) [anchor=0] {\small A};
\node at (axis cs:0.90, 0.85) [anchor=270] {\small B};
\end{axis}
\end{tikzpicture}